\documentclass[twocolumn]{emulateapj}

\usepackage[usenames,dvipsnames,svgnames,table]{xcolor}
\usepackage{graphicx,color,soul}
\usepackage{latexsym}
\usepackage{amsmath,amssymb}      
\usepackage[draft=false]{hyperref}
\hypersetup{
     colorlinks   = true,
     citecolor    = blue
}

\def\apj{ApJ}
\def\apjl{ApJ}
\def\apjs{ApJS}
\def\aap{A\&A}
\def\jcap{J. Cosmology Astropart. Phys.}
\def\mnras{MNRAS}
\def\na{New A}
\def\nar{New A Rev.}
\def\prd{Phys.~Rev.~D}
\def\nat{Nature}

\newcommand{\msun}{\mbox{$M_\odot$}}
\newcommand{\rsun}{\mbox{$R_\odot$}}

\begin{document}

\title{Relativistic Bondi-Hoyle-Lyttleton Accretion onto a Rotating Black-Hole: \\Density Gradients}

\author{F. D. Lora-Clavijo}
\affiliation{Instituto de Astronom\'{\i}a, Universidad Nacional Aut\'{o}noma de M\'{e}xico, AP 70-264, Distrito Federal 04510, M\'{e}xico.\\
Grupo de Investigaci\'on en Relatividad y Gravitaci\'on,
Escuela de F\'isica, Universidad Industrial de Santander,
\\A. A. 678, Bucaramanga 680002, Colombia.}

\author{A. Cruz-Osorio}
\affiliation{Instituto de Astronom\'{\i}a, Universidad Nacional Aut\'{o}noma de M\'{e}xico, AP 70-264, Distrito Federal 04510, M\'{e}xico.}

\author{Enrique Moreno M\'endez}
\affiliation{Instituto de Astronom\'{\i}a, Universidad Nacional Aut\'{o}noma de M\'{e}xico, AP 70-264, Distrito Federal 04510, M\'{e}xico.\\
Universidad An\'ahuac M\'exico Sur, Av. de las Torres 131, 01780, D.F., M\'éxico.}

\email{FDLC: fdlora@astro.unam.mx; \\ ACO: aosorio@astro.unam.mx; \\EMM: enriquemm@ciencias.unam.mx}

\date{\today}

\begin{abstract}
In this work, we present, for the first time, a numerical study of the Bondi-Hoyle accretion with density gradients in the fully relativistic regime. 
In this context, we consider accretion onto a Kerr Black Hole (BH) of a supersonic ideal gas, which has density gradients perpendicular to the relative motion. 
The set of parameters of interest in this study are the Mach number, ${\cal{M}}$, the spin of the BH, $a$, and the density-gradient parameter of the gas, $\epsilon_\rho$.  
We show that, unlike in the Newtonian case, all the studied cases, especially those with density gradient, approach a stationary flow pattern. 
To illustrate that the system reaches steady state we calculate the mass and angular momentum accretion rates on a spherical surface located almost at the event horizon. 
In the particular case of ${\cal{M}} = 1$, $\epsilon_\rho = 0.5$ and BH spin $a = 0.5$, we observe a disk-like configuration surrounding the BH.
Finally, we present the gas morphology and some of its properties. 
\end{abstract}


\maketitle

\section{Introduction}
\subsection{Accretion Onto Astrophysical Black Holes}
{\it Stellar-Mass Black Holes (sBHs):}  
Astrophysical black holes (BHs) come in a variety of masses and spins.  sBH candidates with  $4 \msun \lesssim M_{BH} \lesssim 30 \msun$ have been observed \citep[see, e.g., table 1 in][and references therein]{2013arXiv1305.5543M}.
The masses of the BHs can be obtained through two main channels.  The first of which is by collapse of the progenitor star and capture of fallback mass.  The second channel is through mass transfer from a companion star.  Of course, other channels are possible; e.g., the BH could be the merger of two compact objects (NS-NS, BH-NS, WD-WD, NS-WD, BH-WD, etc.), a compact object and a main sequence (MS) star or through a common-envelope phase; the last two mechanisms could produce a Thorne-\.Zytkow Objects (T\.ZO) that eventually forms a BH \citep{1975ApJ...199L..19T}.

Besides the masses of sBHs it has also been possible to obtain their spins.  
These may reveal important features of the BHs history.
Some of these spins\footnote{The spin, or Kerr parameter, is defined as $a_\star = \frac{a}{M_{BH}} = \frac{cJ_{BH}}{GM_{BH}^2}$ (where $J_{BH}$ is the angular momentum of the BH and $M_{BH}$ is its mass) and it varies between $-1$, for accretion disks which counterrotate with respect to the BH, to $0$ for nonspinning BHs, to $+1$ for corrotating accretion disks.} have been estimated through three different methods and seem to be distributed between $a_\star \sim -0.2$ and $\gtrsim 0.98$.  
All these methods depend on observing phenomena of an accretion disk surrounding the sBH, thus a donor star is necessary.
One method for measuring sBH spins is by fitting the X-ray continuum.
\citep[see][for a review]{2015pabh.book..295M}.  
A second method is through X-ray reflection spectroscopy \citep[of the Fe K-$\alpha$ line; see][for a review]{2015pabh.book..277R}.  
The third method that has been used is through Quasi-Periodic Oscillations (QPOs) from the precession of the accretion disk \citep[see, e.g.,][]{2005AA...438..999A}.

Stellar-mass BHs may acquire their spins through a variety of mechanisms \citep{2002ApJ...575..996L,2005A&A...443..643Y,2006ApJ...637..914W}.  
In most BHs known in Low-Mass X-ray Binaries (LMXBs), the spins may be the result of pre-BH formation processes, either by accretion through mass transfer \citep{2011ApJ...727...29M}, or through tidal-synchronization after a post-Case-C-mass-transfer-and-common-envelope episode \citep{2007ApJ...671L..41B,2014ApJ...781....3M}; this mechanism may even lead to the production of gamma-ray burst and hypernova events (e.g., \citealt{2008ApJ...685.1063B,2014arXiv1411.7377M}; but see also
\citealt{2014arXiv1408.2661F}).

For the BHs in High-Mass X-ray Binaries (HMXBs), the spin estimates seem to pile on the high end, i.e., they vary between 0.84 and $>0.98$.
Stellar evolution in binaries has trouble explaining said spins and so do core-collapse mechanisms (Moreno M\'endez \& Cantiello, in progress), thus \citet{2008ApJ...689L...9M,2011MNRAS.413..183M} have suggested that wind-driven mass transfer with hypercritical accretion may be necessary to explain these binaries.

{\it Intermediate-Mass Black Holes (IMBHs):}
IMBHs are the logical intermediary between sBHs and supermassive BHs (SMBHs).  
If the later are produced by accretion onto the former or from mergers of massive stars or sBHs, then IMBHs should be abundant.  
On the other hand, if they are only produced from core-collapse of Population III stars, they may have already turned into SMBHs.
Indeed, they are excellent candidates to explain the observed Ultra Luminous X-ray (ULX) sources.  
Nonetheless, many ULXs have been identified with LMXBs \citep{2014Natur.514..202B} and HMXBs \citep{2014Natur.514..198M,2013Natur.503..500L}.  
There seems to be, however, a good candidate of an IMBH in a ULX where, using High-frequency QPOs (with 3:2 ratio), \citet{2014Natur.513...74P} use inverse-scaling of sBH as well as a relativistic precession model to determine a mass of $M_{BH} \simeq 400 \msun$.

{\it Super-Massive Black Holes (SMBHs):}
Explaining the mass of supermassive black holes (SMBHs) is one of the most interesting current problems of astrophysics. 
A standard approach to the problem assumes that these holes are the result of the accretion onto intermediate-mass black-hole (IMBH) seeds. 
This raises the question of where such seeds come from in the first place.
The answer to this question 
has opened a wide set of possibilities. 
It could be that disks formed in protogalaxies allow the infall of matter that collapses to form black holes of $10^5 M_{\odot}$ \citep{2004MNRAS.354..292K}. Alternatively, seeds of $10^3 M_{\odot}$ can be formed due to the runaway collision of compact stellar clusters in low metallicity protogalaxies at $z \sim 10-20$ that additionally allow accretion during the quasar era  \citep{2009ApJ...694..302D}. 
Also, seeds can be the result of the core collapse of dense clusters and form $10^5M_{\odot}$ black holes \citep{2011ApJ...728...98D}. 
Another alternative is that seeds of $10^5M_{\odot}$ can be formed due to the collapse of very massive stars \citep{2009JCAP...08..024U} at extremely low metallicity.
A more detailed model in this direction proposes that a supermassive primordial star forms in a region of the Universe with a high molecule-dissociating background-radiation field, and collapses directly into a $10^4 - 10^6 M_{\odot}$ black-hole seed \citep{2013ApJ...771..116J}. 
More recently, it has been proposed that primordial black holes, during the radiation-dominated era, can grow up to $10^3-10^6 M_{\odot}$  \citep{2013JCAP...12..015L}.

Based on the study of the evolution of phase space distributions, the growth process of seeds in standard analysis of SMBHs considers that these are primarily fed by collisionless dark
matter or stars \citep{1977ApJ...211..244L,2002NewA....7..385Z}. 
Previous results show that the timescale is extremely long for collisionless dark matter to contribute significantly to black-hole growth  \citep[for instance][]{2003MNRAS.339..949R}. 
In \cite{2011MNRAS.415..225G,2011MNRAS.416.3083G} the conditions for stable accretion and runaway accretion of an ideal gas have been studied.
Other more generalized cases, consider non-radial accretion processes which require the particles to overcome the angular momentum barrier in order to get the gas to the center of a galaxy \citep{2006MNRAS.373L..90K}. 

The assumption of self-interacting dark matter (SIDM) has been used to analyze the problem of SMBH growth. 
For instance in \citet{2002ApJ...568..475B} and \citet{2002PhRvL..88j1301B} SIDM is used to model direct dynamical collapse and SMBH formation due to the gravothermal catastrophe.  
Furthermore, this would explain the different SMBH seed masses in terms of the redshift at which the collapse took place.  
\cite{2005A&A...436..805M} show that fermionic dark matter can feed SMBH seeds to make them grow up to $10^3 -10^6 M_{\odot}$. 
Instead, \cite{2008MNRAS.391.1403S} propose that the self-interaction is introduced using a polytropic equation of state for the pressure, thus, finding another mechanism for SMBH formation.  
\cite{2000PhRvL..84.5258O} \citep[and later][]{2006MNRAS.365..345H} analyzed black hole formation due to the collapse of SIDM and also studied SMBH formation and growth due to the collapse and accretion of SIDM.  
\cite{2014MNRAS.443.2242L}  studied the accretion of a SIDM into a SMBH for the particular case of radial flows. 
They considered the evolution of space-time in order to have the formally correct black-hole growth rate.

The characterization of SMBH spin is of vital importance since it probes their growth history as well as their formation. 
In essence, scenarios in which SMBH growth is dominated by BH-BH mergers predict a population of modestly spinning SMBHs, whereas growth
via gas accretion can lead to either, a rapidly-spinning, or a very slowly-spinning population \citep{1996MNRAS.283..854M,2005ApJ...620...69V,2015pabh.book..277R}. SMBH spin can also be a potent energy source, and may well drive the powerful relativistic jets that are seen from many black hole systems \citep[e.g.,][]{1977MNRAS.179..433B}.

\subsection{Bondi-Hoyle-Lyttleton Accretion}

Bondi-Hoyle-Lyttleton (BHL) accretion deals with the evolution of a homogeneously distributed gas moving uniformly toward a central compact object \citep{1939PCPS...35..405H,1944MNRAS.104..273B}. 
Depending on whether the velocity of the gas is supersonic (or not) a shock cone is formed (or not). 
This process shows interesting properties when considered within the Newtonian and relativistic regimes, which have been explored based on several numerical studies. 
In the classical regime, which is ruled by Newtonian gravity, the most important subjects are the consequences on the morphology of the wind and the supersonic shocks that develop. A summary of results, under Newtonian gravity, can be found in 
\citep{2004NewAR..48..843E,2005A&A...435..397F}.
 
Unlike in the Newtonian regime, the relativistic approach allows the study of BHL accretion in regions where the gravitational field is strong. 
Some studies in this direction have been carried out. 
The first one, performed by \cite{1989ApJ...336..313P}, studied the different accretion patterns developed by the relativistic gas during the accretion onto a BH. 
Later on, considering axial and equatorial symmetries, \cite{1998ApJ...494..297F,1998MNRAS.298..835F,1998ApJ...507L..67F,1999MNRAS.305..920F} reviewed the results obtained by \cite{1989ApJ...336..313P} using more accurate methods. 
In the astrophysical context and using equatorial symmetry, \cite{2011MNRAS.412.1659D} showed that the shock cone vibrations can be associated with sources of QPOs. 
They also found a flip-flop type of unstable oscillation of the shock cone. However, it was later shown by \cite{2012MNRAS.426..732C} that the flip-flop oscillation of the shock cone depends on the coordinates used to describe the rotating black hole, specifically, it was found that the flip-flop oscillation does not appear when Kerr-Schild coordinates are used to describe the rotating black hole. 
More recently, considering axially symmetric fluxes, in \cite{2013MNRAS.429.3144L}, the shock cone oscillations as a potential source of low and high frequency QPOs were studied. 
The ultrarelativistic BHL accretion onto a rotating black hole was recently reported considering axisymmetric fluxes \citep{2013MNRAS.428.2171P} and   considering equatorial symmetry \citep{2013AIPC.1548..323C}. 
More realistic scenarios introduce astrophysically relevant ingredients like magnetic fields \citep{2011MNRAS.414.1467P}, radiative terms \citep{2011MNRAS.417.2899Z} and full general relativity in the context of supermassive black hole binary mergers \citep{2010PhRvD..81h4008F}.

This work uses penetrating coordinates (Kerr-Schild), which offer the possibily to place the excision inside the event horizon of the BH.
It is well known that what happens inside a BH, stays inside the BH, thus this further avoids the implementation of boundary conditions there.

In the Newtonian regime, the BHL problem is attended by assuming the accretor is point-like and the size of the accretor is determined in terms of the accretion radius. In our case we are doing the exact opposite; this implies that if we want to resolve the accretor, i.e. the Schwarzschild radius must be similar in order of magnitud to the accretion radius, thus, the velocity of the wind must be high. 
This is so that the ratio of the accretion radius (as well as that of the compact object) can be small enough as compared to the exterior numerical domain radius (the ratio must be at least an order of magnitude) so that no numerical artifacts affect the simulations \cite{2012MNRAS.426..732C,1998ApJ...507L..67F}. Hence, this allows to have good resolution and a reasonable computational timescale.   The accretion radius is defined to help decide when a particle falls into the compact object.  Such a scale is determined by the velocity of the wind and the equation of state of the gas. In our case we use the accretion radius only to choose the exterior numerical domain, that is about ten times the accretion radius.

Another problem one may face is that the relativistic Euler equations may diverge or develop unphysical results; thus an atmosphere is implemented which prevents the rest mass density to be small enough such as to provoke said effects.
The atmosphere we use is $\rho=max(\rho,10^{-14})$; such a value allows the convergence of our numerical methods. The need
of the atmosphere is in fact one of the reasons why it is not
-at the moment- trivial to simulate the evolution of fluids with
ultralow density.
Given these numerical limitations, the maximum density difference between the BH\footnote{Here we understand by BH density the BH mass divided by the spherical volume using the radius of its event horizon, thus $\rho_{BH} \propto M_{BH}^{-2}$.  In this paper we are working under the assumption that the space-time background is unaffected by the fluid contribution, i.e., we have a test fluid, which is consistent with our astrophysical models.} and the non-homogeneous wind is of order $\rho_{BH}/\rho_{w} \simeq 10^{12}$.  
Also, the velocities in our models will be restricted to values between 0.1$c$ and 0.5$c$.   
This minimizes the possible scope of astrophysical scenarios that can be dealt with.

\subsection{Outlook}

In this manuscript we present for the first time a numerical model of relativistic BHL accretion onto a Kerr BH of a non-homogeneous gas cloud. This problem was addressed initially, in the Newtonian regime, by \cite{1995A&A...295..108R} and \cite{1997A&A...317..793R,1999A&A...346..861R}. 
They investigated the hydrodynamics of three-dimensional classical BHL accretion; especially the accretion of angular momentum from a non-homogeneous medium and discussed some consequences for models of wind-fed X-ray sources. 
They found that the models with a density gradient exhibit non-stationary flow patterns, even though the Mach cone remains fairly stable. 
In more recent work, also in the Newtonian regime, \citet{2014arXiv1410.5421M} model the orbital inspiral of a neutron star (NS) through the envelope of its giant-branch companion during a common envelope (CE) episode. They found that the presence of a density gradient strongly limits the accretion by imposing a net angular momentum to the flow around the NS.

Based on the limitations stated above, we accordingly list which models are appropriate for our different simulations which have been left unitless in order to maximize the posible scenarios in which they would be useful.  
Hence, in what follows, the mass of the BH shall be denoted $M$. Following the usual convention, we shall mostly use units in which
$G = c = 1$. Thus, length and time are measured in units of 
$M$ with $1~\msun~\equiv~1.48~\times~10^{5}$ cm $\equiv~4.93~\times~10^{-6}$ s.

This paper is organized as follows:  Section~\ref{sec:models} proposes models for BHs of different masses.  Section~\ref{sec:equations} describes  the relativistic hydrodynamic equations as well as the numerical methods employed by our code.  Section~\ref{sec:results} shows our numerical results. And in section~\ref{sec:conclusions} we discuss them and produce our conclusions.

\section{Models}
\label{sec:models}

We have produced a set of models which are BH mass independent, therefore, once the mass of the BH has been chosen, its density is established and one may obtain the wind density that hits the BH.  
The velocities of the flows hitting the BH are 0.1c, 0.2c, 0.3c, 0.4c and 0.5c.

We next list a set of astrophysical situations where our model would become a good fit with the proper scaling.

\subsection{For Stellar-Mass BHs}

{\it Relativistic winds and shockwaves:}
An astrophysical scenario where our models may apply for stellar-mass BHs is during a hypernova in a binary system.
Suppose we have a short-orbital-period binary ( $P \lesssim 0.4$ day) with a massive WR star ($M_{WR} > 10\msun$) and a $\sim 10 \msun$ BH (e.g., Cyg X-3 may be a system in a situation similar to this description).  
The WR tidally synchronizes and rotates rapidly.
Such a star will collapse into a compact object and will likely launch a GRB/HN explosion \citep[see, e.g.,][]{2011ApJ...727...29M}.
If $\sim 5 \msun$ collapse, the resulting BH could have $a_\star \gtrsim 0.8$ and the available energy would be $\gtrsim 10^{54}$ erg (or 1000 Bethes; 1 Bethe $= 1 $ B $ = 10^{51}$ erg).
If $\sim 5 \msun$ of material are expelled with a fraction (say $10\%$ to $50\%$ of this energy)
their average velocity would be between $0.1 c$ and $0.7 c$.
At the time of the explosion, the orbital separation is about 3 $\rsun$ to 4 $\rsun$.
Thus, the density of the ejecta at the orbit of the BH could be as large as 
one fifth of its density inside the star if the expansion is mostly equatorial or a hundredth if the expansion is spherically symmetric. 
This translates to densities as high as $10^3$ g cm$^{-3}$ if material from the C core is expelled.
For a $10 \msun$ BH,  the density is $\rho_{BH} \lesssim 10^{16}$ g cm$^{-3}$.  
Thus our model deals with $\rho_{w} \simeq 10^4$ g cm$^{-3}$, which is 1 to 2 orders of magnitude above what this model provides.  
Hence, our simulations give a good qualitative description of what should occur for such a scenario.

\subsection{IMBHs:  }

IMBHs have a density range of $\rho_{BH} \sim 10^{13}$ g cm$^{-3}$ ($M_{BH} = 10^2 \msun$) to $\rho_{BH} \sim 10^{5}$ g cm$^{-3}$  ($M_{BH} = 10^6 \msun$).  
Thus, given that our simulations are bound to $\rho /\rho_{BH} \sim 10^{-12}$, we have a wind density regime that goes from $10$ g cm$^{-3}$ to $10^{-7}$ g cm$^{-3}$.

A scenario similar to the previous one would be ideal.
I.e., an IMBH in the mass range of $10^2~\msun~\lesssim~M_{BH}~\lesssim~10^4~\msun$, located, e.g., near the center of a globular cluster which were to tightly capture a massive WR star (maybe through a common envelope phase).
If the orbital period was, again, 0.4 days, the orbital separation, $d$, would be $11~\rsun~\lesssim~d~\lesssim~50~\rsun$, respectively.  
The WR radius being some 50 to 10\% of its Roche lobe (RL) radius (the BH radius being less than about a thousandth of its RL radius).
For such an orbital separation the  density of the ejecta should be around $5~\times~10^{-2}$  g cm$^{-3}$  to $10^{-3}$  g cm$^{-3}$, at $11~\rsun$; and $10^{-3}$ g cm$^{-3}$ to $5~\times~10^{-4}$ g cm$^{-3}$, at $50~\rsun$. 
Thus, these models are well within the range of our numerical calculations.

\subsection{SMBHs:  }

SMBHs \citep[$10^6 \msun$ to $10^9 \msun$; although there is recent evidence for a SMBH as large as $M_{BH} \gtrsim 10^{11} \msun$,][]{2014ApJ...795L..31L} have a density range that goes from $10^5$ g cm$^{-3}$ to $10^{-1}$ g cm$^{-3}$.
The wind densities in our numerical simulations are, hence, confined between $10^{-7}$ g cm$^{-3}$ and $10^{-13}$ g cm$^{-3}$.

Among the possible scenarios where our models may be applied for SMBHs are, e.g., a SMBH binary where one of the BHs has relativistic jets pointing along the orbital plane.  
As the second BH steps into or out of the jet of the first BH, it is submerged in a stream of material with a density gradient.

It is conceivable that the last model discussed for IMBHs (a nearby GRB pointing towards the BH) could similarly apply for SMBHs but at somewhat larger distances.

\section{Relativistic Hydrodynamic Equations and numerical methods}
\label{sec:equations}
\subsection{Relativistic Hydrodynamics Equations}

In order to solve numerically the relativistic Euler equations, we use the 3+1 decomposition of space-time, in which the space-time is foliated with a set of non-intersecting space-like hypersurfaces  $\Sigma_t$  
\citep[see e.g.][]{2008itnr.book.....A,2010nure.book.....B,2013rehy.book.....R}. 
The space-time is described with the line element

\begin{equation}
ds^2 = -\alpha^2dt^2 + \tilde{\gamma}_{ij}(dx^i + \beta^i dt)(dx^j + \beta^j dt) , 
\label{eq:lineelement}
\end{equation}

\noindent where $\alpha$ is the lapse function, $\beta^i$  are the shift vector components and $\tilde{\gamma}_{ij}$ are the components of the induced three metric that relates proper distances on the spatial hypersurfaces. 

The background space-time corresponds to a rotating black hole in Kerr-Schild coordinates, which allow us to place the inner boundary of the computational domain inside the event horizon. A discussion of the advantage on the use of these coordinates can be found in \citet{2012MNRAS.426..732C}. Once the geometrical elements of the space-time background are known, it is necessary to track the evolution of the fluid, for which we write down the general relativistic Euler equations. For a generic space-time these can be derived from the local conservation of the stress-energy tensor 

\begin{equation}
\nabla_{\nu} (T^{\mu \nu}) = 0,
\label{eq:Con1}
\end{equation}

\noindent  and the local conservation of the rest mass density 

\begin{equation}
\nabla_{\nu} (\rho u^{\nu}) = 0, 
\label{eq:Con2}
\end{equation}

\noindent where $\rho$ is the proper rest mass density, $u^{\mu}$ is the four-velocity of the fluid and $\nabla_{\nu}$ is the covariant derivative consistent with the four-metric $g_{\mu \nu}$ of the space-time (\ref{eq:lineelement}). 

We assume the matter field in the above equations is that of a perfect fluid with stress-energy tensor 

\begin{equation}
T_{\mu \nu} = \rho hu^{\mu} u^{\nu} + pg^{\mu \nu}, 
\end{equation}

\noindent where $p$ is the pressure, $g_{\mu \nu}$ are the components of the four-metric and $h$ the relativistic specific enthalpy given by $h = 1 + \epsilon + p/\rho$, where $\epsilon$ is the rest frame specific internal energy density of the fluid.

It is well known that Euler's equations develop discontinuities in the hydrodynamical variables even if smooth initial data are considered. Thus one may solve hydrodynamics equations using finite volume methods, as long as the system of equations is written in a flux balance law form, which in turn requires the definition of conservative variables.

In order to obtain the general relativistic Euler equations as a set of flux balance laws \citep{1997ApJ...476..221B,2000PhRvD..61d4011F}, we project the local conservation equations  along the space-like hypersurfaces and the normal direction to such hypersurfaces. A straightforward  calculation yields the set of equations in the desired form

\begin{equation}
\frac{1}{\sqrt{-g}}\left[ \partial_{t} (\sqrt{\tilde{\gamma}} {\bf q} ) + \partial_{i} \left(\sqrt{-g}{\bf f}^{(i)}({\bf q})\right) \right]= {\bf s}({\bf q}), 
\label{eq:HydroEvolve}
\end{equation}

\noindent where $g$ is the determinant of the four-metric (\ref{eq:lineelement}), ${\bf q}$ is a vector of conservative variables, ${\bf f}^{(i)}({\bf q})$ are the fluxes along each spatial direction and ${\bf s}({\bf q})$ is a source vector. These last quantities are given by: 

\begin{widetext}
\begin{eqnarray}
{\bf q} &=& [D, ~~ M_j, ~~ \tau ]^T = [\rho \Gamma, ~~ \rho h \Gamma^2 v_j,  ~~ \rho h \Gamma^2 -p - \rho \Gamma ]^T, \\
{\bf f}^{(i)}({\bf q}) &=& \left[\left(v^i - \frac{\beta^i}{\alpha}\right)D, ~~ \left(v^i - \frac{\beta^i}{\alpha} \right)M_j + \delta_{j}^{i}p, ~~ \left(v^i - \frac{\beta^i}{\alpha} \right)\tau + v^i p\right]^T, \\
{\bf s}({\bf q}) &=& [ 0, ~~ T^{\mu \nu} g_{\nu \sigma} \Gamma^{\sigma}_{\mu j}, ~~ T^{\mu 0} \partial_{\mu}\alpha - \alpha T^{\mu \nu} \Gamma^{0}_{\mu \nu}]^T.\label{eq:sOFq}
\end{eqnarray}
\end{widetext}

\noindent In these expressions,  $\tilde{\gamma}=det(\tilde{\gamma}_{ij})$ is the determinant of the spatial metric, $\Gamma^{\sigma}{}_{\mu \nu}$ are the Christoffel symbols and  $v^{i}$ is the 3-velocity measured  by an Eulerian observer and defined in terms of the spatial part of the 4-velocity $u^{i}$, as $v^{i}=u^{i}/\Gamma + \beta^{i}/\alpha$, where $\Gamma$ is the Lorentz factor given by $\Gamma=1/\sqrt{1-\tilde{\gamma}_{ij} v^{i}v^{j}}$.

It is still necessary to close the system of equations (\ref{eq:HydroEvolve}), for which an equation of state relating $p=p(\rho,\epsilon)$ is used. We choose the gas to obey an ideal gas equation of state 

\begin{equation}
p=\rho \epsilon(\gamma -1),
\label{eq:EoS}
\end{equation}

\noindent  where $\gamma$ is the adiabatic index or the ratio of specific heats. Something to stand out is that the relativistic sound velocity $c_{s}$ for an ideal equation of state can be written as $c_{s}^{2} = p\gamma(\gamma-1)/[p\gamma - \rho (\gamma -1)]$, where its asymptotic value or its maximum permitted value is $c_{s_{max}}=\sqrt{\gamma -1}$. Thus, the choice of our initial values is restricted to this condition.

\subsection{Numerical methods}

{\it Gas Evolution:} The general relativistic Euler system of equations (\ref{eq:HydroEvolve}), is solved in time using the method of lines, that uses a third order total variation diminishing (TVD) Runge-Kutta time integrator \citep{Shu1988439}. 
These are discretized using a finite volume approximation together with high resolution shock capturing methods. Specifically, we use the HLLE \citep{doi:10.1137/1025002,doi:10.1137/0725021} approximate Riemann solvers in combination with the minmod linear piecewise reconstructor. 
The numerical fluxes and sources in (\ref{eq:HydroEvolve}) depend both on the conservative and on the primitive variables ${\bf w}=(\rho_0,v^i,p)$. Then, in order to express primitive in terms of conservative variables, we use a Newton-Raphson algorithm each time step within the evolution scheme.

{\it Boundary Conditions:} We study numerically the relativistic gas on the equatorial plane, in the domain $ [r_{exc},r_{max}] \times  [0,2\pi)$ with resolution $(\Delta r, \Delta \phi ) = (0.158, 0.05)$ for all cases. We choose the interior boundary $r_{exc}$ to be inside the black hole horizon, were we apply a numerical excision \citep{1992PhRvL..69.1845S}, i.e., we apply a cutoff inside the event horizon which is possible due to the Kerr-Schild coordinates used by CAFE. The exterior boundary $r_{max}$, is splitted in to two halves, one in which the gas enters the domain where we apply inflow boundary conditions, and a second half where the gas leaves the domain and we apply outflow boundary conditions there. Besides, in the angular domain we use periodic boundary conditions.

\begin{figure*}
\begin{center} 
 \begin{tabular}{cc} 
 \includegraphics[width=9.5cm]{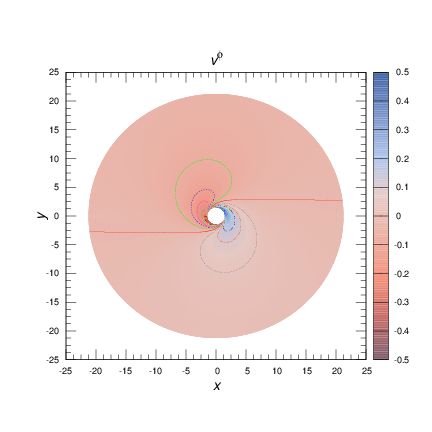} &\hspace{-1.2cm}
 \includegraphics[width=9.5cm]{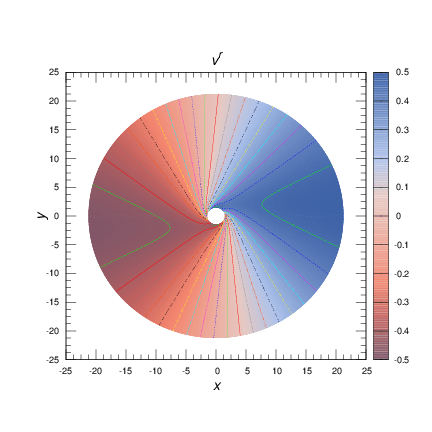}\\
  \end{tabular}
  \end{center}
  \caption{\label{fig:vel} This figure shows the components of the velocity field at the initial time.  The parameters used to illustrate this are:  BH spin, $a = 0.9$; the initial wind velocity at infinity, $v_\infty = 0.5 c$, and the sound speed at infinity is $c_{s\infty}$.  We remind the reader that length is in units of $M$ with $1~\msun~\equiv~1.48~\times~10^{5}$ cm.}
  \end{figure*}

{\it Initial data:} As initial data, we consider a wind, that fills the whole domain, moving on the equatorial plane along the $x$ direction with non constant density profile. We characterize the initial velocity field $v^i$ in terms of the asymptotic initial velocity $v_{\infty}$ as done in \citet{2012MNRAS.426..732C,1999MNRAS.305..920F}, where the relation $v^2=v_iv^i=v^2_{\infty}$ is satisfied.
Using Kerr-Schild coordinates the explicit expressions for the velocity vector field $v^i$ are given by:
\begin{eqnarray}
v^{r}&=&H_{1}v_{\infty} \cos \phi+H_{2}v_{\infty} \sin \phi, \\
v^{\phi}&=&-H_{3}v_{\infty} \sin \phi +H_{4}v_{\infty} \cos \phi.
\end{eqnarray}
where $H_i$ ($i = 1, 2, 3, 4$) represent functions associated with the components of the three metric.
\begin{eqnarray}
H_{1}&=&\frac{1}{\sqrt{\tilde{\gamma}_{rr}}},\\
H_{2}&=&\frac{ H_{3}H_{4}\tilde{\gamma}_{\phi\phi}+H_{1}H_{3}\tilde{\gamma}_{r\phi}}{H_{1}\tilde{\gamma}_{rr}+H_{4}\tilde{\gamma}_{r\phi} },\\
H_{3}&=&\frac{H_{1} \tilde{\gamma}_{rr}+H_{4}\tilde{\gamma}_{r\phi}}{\sqrt{(\tilde{\gamma}_{rr}\tilde{\gamma}_{\phi\phi}-\tilde{\gamma}_{r\phi}^{2})( H_{1}^{2}\tilde{\gamma}_{rr} +H_{4}^{2}\tilde{\gamma}_{\phi\phi} + 2H_{1} H_{4}\tilde{\gamma}_{r\phi})} },\\
H_{4}&=&-\frac{2\tilde{\gamma}_{r\phi}}{\sqrt{\tilde{\gamma}_{rr}}\tilde{\gamma}_{\phi\phi}}.
\end{eqnarray}
The profiles of the components of the field velocity can be seen in figure~\ref{fig:vel}.

Following Ruffert's work \citep{1995A&A...295..108R,1997A&A...317..793R,1999A&A...346..861R}, the initial density gradient is chosen in such way that it is perpendicular to the gas motion and also proposed as an hyperbolic function in order to serve as a cutoff at large distances for large density gradients. The density distribution, in polar coordinates, is given by the following expresion 

\begin{eqnarray}
\rho_{ini} = \rho_{0} \Big\{ 1- \frac{1}{2}\tanh\Big[2\epsilon_{\rho}\frac{(r \sin[\phi] + a \cos[\phi])}{r_{a}}\Big] \Big\}, \label{eq:Gd}
\end{eqnarray}

\noindent where $\rho_{0}$ is a constant density value, $a=J/M$ is the spin of the black hole, $\epsilon_{\rho}$ is the parameter specifying the magnitude of the density gradient and $r_a$ is the accretion radius, which is defined in terms of the asymptotic values of the sound speed $c_{s_\infty}$ as \cite{1989ApJ...336..313P}

\begin{equation}
r_a = \frac{M}{v_{\infty}^2 + c_{s \infty}^2}.
\end{equation}

\begin{figure*}
 \begin{center}
 \begin{tabular}{cc} 
 \includegraphics[width=9.5cm]{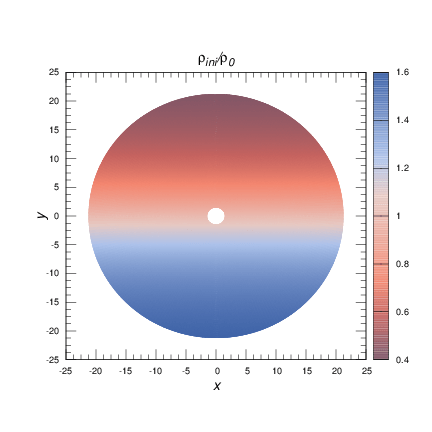}&\hspace{-1.2cm}
\includegraphics[width=9.5cm]{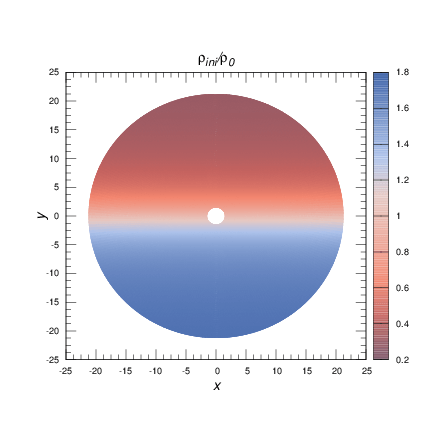}\\
  \end{tabular}
  \end{center} 
  \caption{\label{fig:dens} This figure shows the initial density profile for two values of the density gradient parameter, $\epsilon_\rho = 0.2$ and $\epsilon_\rho = 0.5$.  As can be observed, for higher values of $\epsilon_\rho$ the drop in density is considerably more pronounced.  We remind the reader that length is in units of $M$ with $1~\msun~\equiv~1.48~\times~10^{5}$ cm.}
  \end{figure*}

Figure~\ref{fig:dens} illustrates the density profile for two values of $\epsilon_\rho$.

Once we fix the value of $c_{s \infty}$ and assume the gradient density profile (\ref{eq:Gd}), the pressure can be found from the equation of state as: $p_{ini} = c_{s \infty}^2 \rho_{ini}/(\gamma - c_{s \infty}^2 \gamma_1)$, where $\gamma_1=\gamma/(\gamma -1 )$. In order to avoid negative and zero values on the pressure, the condition $c_{s\infty} < \sqrt{\gamma - 1}$ has to be satisfied. 

Now, $v_{\infty}$ and $c_{s}{}_{\infty}$ are two useful parameters which define the relativistic Mach number at infinity, ${\cal M}^{R}_{\infty}=\Gamma v_{\infty} / \Gamma_s c_{s}{}_{\infty} = \Gamma {\cal M}_{\infty}/\Gamma_s$.  Here $\Gamma_s$ is the Lorentz factor calculated with the speed of sound and ${\cal M}_{\infty}$ is the asymptotic Newtonian Mach number used to parametrize the initial configurations. 
The initial data is parametrized using the Mach number given in Table~\ref{tab:params}.

We use CAFE \citep{2014arXiv1408.5846L}, a fully three-dimensional relativistic-magnetohydrodynamic (MHD) code.
Although the MHD in CAFE is written for a Minkowsky space-time (it does not use curved space-time), the hydrodynamical (HD) routine does utilize the equatorial \citep{2012MNRAS.426..732C,2013AIPC.1548..323C} and axial \citep{2013MNRAS.429.3144L} symmetries in order to allow simulations using fixed curved space-times.
CAFE also solves the Einstein field equations coupled with relativistic HD in spherical symmetry \citep{2012arXiv1212.1421G,2013JCAP...12..015L,2014MNRAS.443.2242L}. 
The numerical methods available in CAFE include several reconstructors.
For instance MINMOD and MC linear piecewise methods; for higher reconstructors CAFE uses PPM parabolic method and WENO5 polynomial method.
All of these reconstructors are used in combination with HLLE approximate Riemann solver flux formula. 
Concerning relativistic MHD, and in order to preserve magnetic field divergence, CAFE uses flux constraint-transport and divergence cleaning methods.
Numerical tests of the numerical implementation in CAFE can be found, for the relativistic HD, in \cite{2013arXiv1303.3999L,2013MNRAS.429.3144L} and for relativistic MHD in \cite{2014arXiv1408.5846L}.

\begin{table}[!h]
\centering
 \begin{tabular}{|c|cc|cc|cc|cc|cc|}\hline\hline
  ${\cal{M}}_{\infty}$ & \multicolumn{2}{c|}{a=0} & \multicolumn{2}{c|}{a=0.3} & \multicolumn{2}{c|}{a=0.5} & \multicolumn{2}{c|}{a=0.7} & \multicolumn{2}{c|}{a=0.9} \\ 
   & $\epsilon_{\rho}$ & $r_{a}$ & $\epsilon_{\rho}$ & $r_{a}$ & $\epsilon_{\rho}$ & $r_{a}$ & $\epsilon_{\rho}$ & $r_{a}$ & $\epsilon_{\rho}$ & $r_{a}$ \\ \hline 
   & $0.0$ & & $0.0$ & & $0.0$ & & $0.0$ & & $0.0$ &       \\ 
 5 & $0.2$ & 3.85 & $0.2$ & 3.85 & $0.2$ & 3.85 & $0.2$ & 3.85 & $0.2$ & 3.85  \\ 
   & $0.5$ & & $0.5$ & & $0.5$ & & $0.5$ & & $0.5$ &       \\ \hline
   & $0.0$ & & $0.0$ & & $0.0$ & & $0.0$ & & $0.0$ &       \\ 
 4 & $0.2$ & 5.88 & $0.2$ & 5.88 & $0.2$ & 5.88 & $0.2$ & 5.88 & $0.2$ & 5.88  \\ 
   & $0.5$ & & $0.5$ & & $0.5$ & & $0.5$ & & $0.5$ &       \\ \hline
   & $0.0$ & & $0.0$ & & $0.0$ & & $0.0$ & & $0.0$ &       \\ 
 3 & $0.2$ & 10.0 & $0.2$ & 10.0 & $0.2$ & 10.0 & $0.2$ & 10.0 & $0.2$ & 10.0  \\ 
   & $0.5$ & & $0.5$ & & $0.5$ & & $0.5$ & & $0.5$ &       \\ \hline 
   & $0.0$ & & $0.0$ & & $0.0$ & & $0.0$ & & $0.0$ &       \\
 2 & $0.2$ & 25.0 & $0.2$ & 25.0 & $0.2$ & 25.0 & $0.2$ & 25.0 & $0.2$ & 25.0  \\ 
   & $0.5$ & & $0.5$ & & $0.5$ & & $0.5$ & & $0.5$ &       \\ \hline \hline
\end{tabular}
\caption{ \label{tab:params} In this table, we summarize the models studied in figures~\ref{MdotVsT} and \ref{PdotVsT}. All the different configurations considered here asumme the relativistic sound speed at infinty is $c_{s \infty} = 0.1$ and adiabatic index $\gamma=5/3$.}
\end{table}

\subsection{Diagnostics}

In order to diagnose the amount of mass and angular momentum accreted by the Kerr BH we implement a detector located as close as possible to the (outer) event horizon at: 
\[
r_+ = M + \sqrt{M^2 - a^2}.
\]
This implies we define a sphere where we compute said scalars.
The following formulae represent both quantities respectively:

\begin{eqnarray}
\dot{M} &=& \int_{0}^{2\pi} \alpha\sqrt{\tilde{\gamma}}D(v^{r} - \beta^{r}/\alpha) d\phi,
\label{eq:MAcc}\\
\dot{P}^{\phi}&=&-\int_{0}^{2\pi} \alpha \sqrt{\tilde{\gamma}} T^{r\phi} d\phi  + \int_{0}^{2\pi}\int_{r_{exc}}^{r_{det}} S^{\phi} drd\phi,
\label{eq:pAcc}
\end{eqnarray}
where $S^\phi$ stands for the $\phi$ component ($j=\phi$) of the source term of equation (\ref{eq:sOFq}), $r_{exc}$ is the excision radius and $r_{det}$ is the radius where the detector lays.

\section{Results}\label{sec:results}
  
\begin{figure*}
\begin{tabular}{cccc} \hline
 ${\cal M}= 2$ & ${\cal M}= 3$ & ${\cal M}= 4$ & ${\cal M}= 5$ \\ \hline
 && \\ 
\multicolumn{4}{c}{$a= 0$}\\ 
\hspace{-4mm} \includegraphics[width=4.7cm,height=4cm]{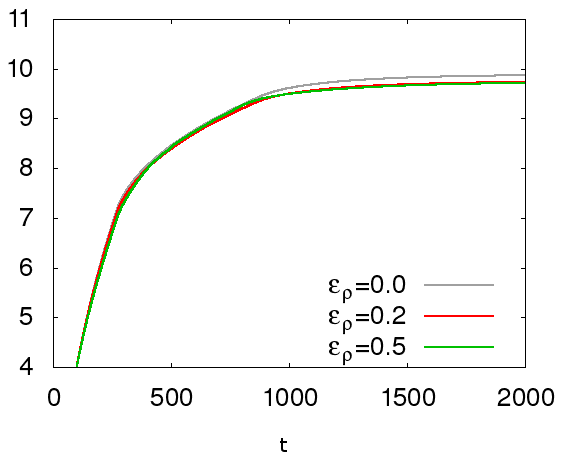}&
\hspace{-5mm} \includegraphics[width=4.7cm,height=4cm]{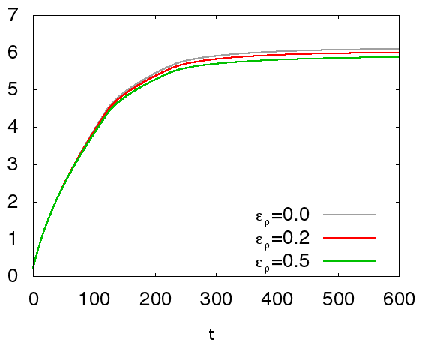}&
\hspace{-5mm} \includegraphics[width=4.7cm,height=4cm]{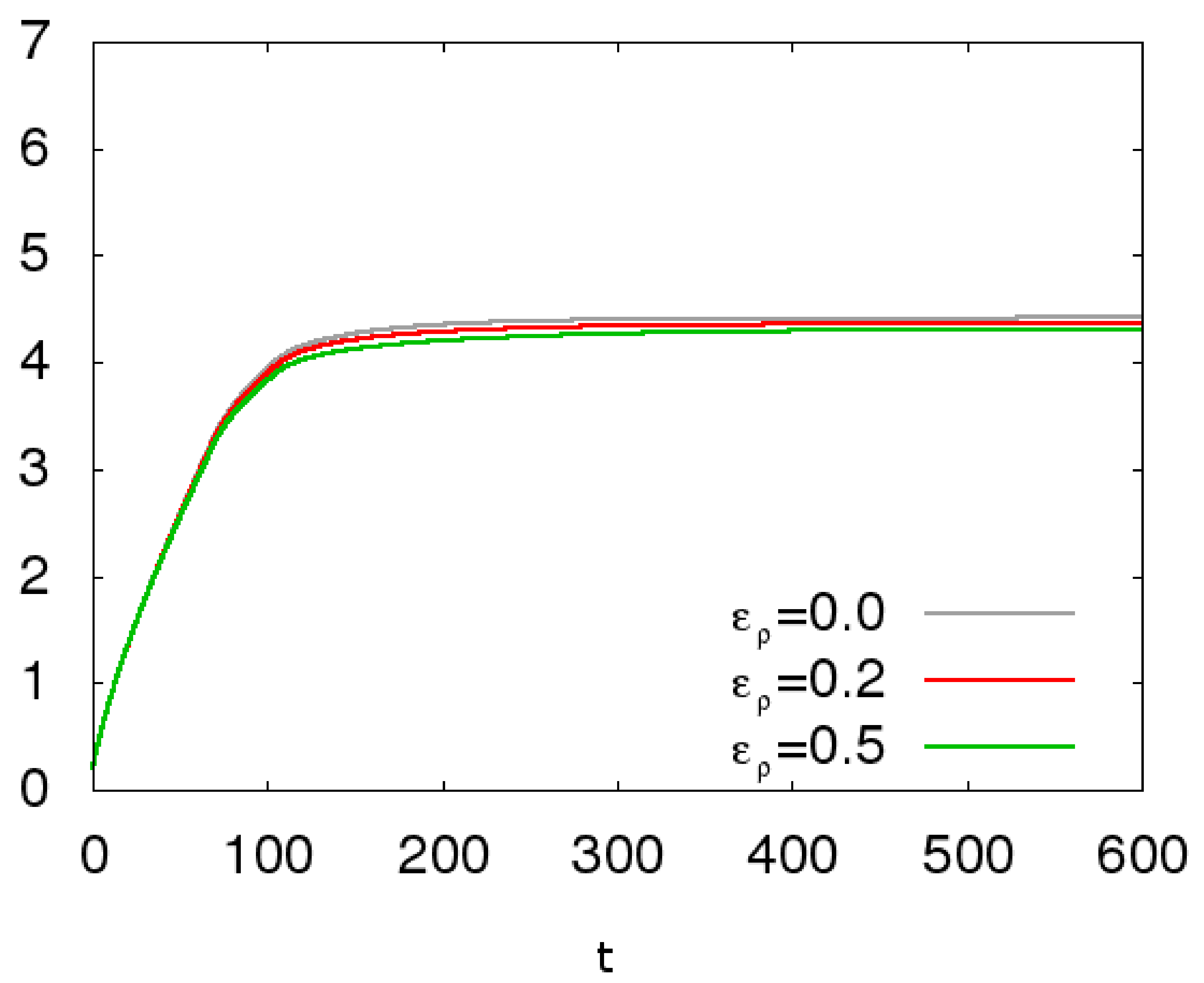}&
\hspace{-5mm} \includegraphics[width=4.7cm,height=4cm]{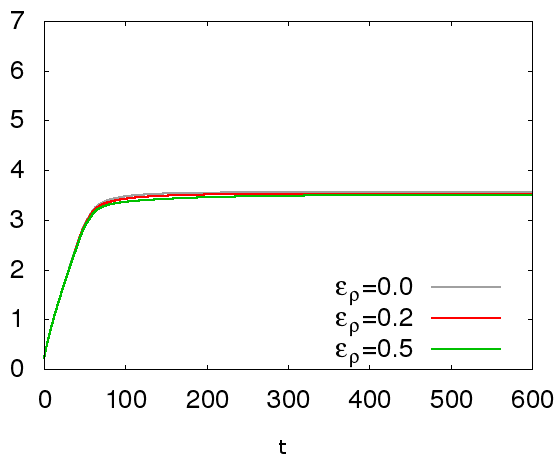}\\
\multicolumn{4}{c}{$a= 0.3$}\\ 
\hspace{-4mm}\includegraphics[width=4.7cm,height=4cm]{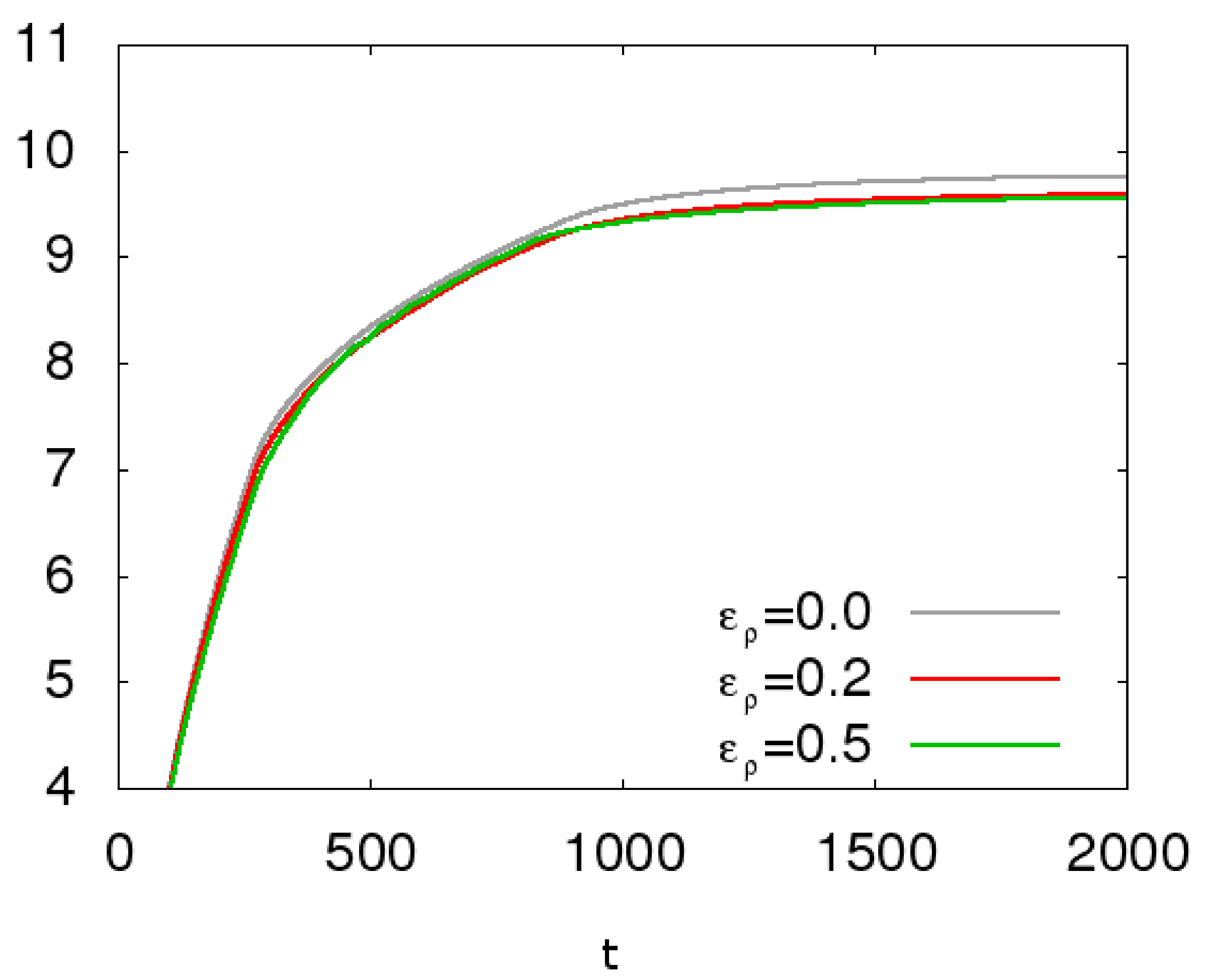}&
\hspace{-5mm}\includegraphics[width=4.7cm,height=4cm]{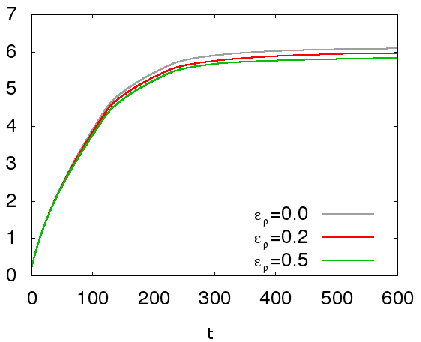}&
\hspace{-5mm}\includegraphics[width=4.7cm,height=4cm]{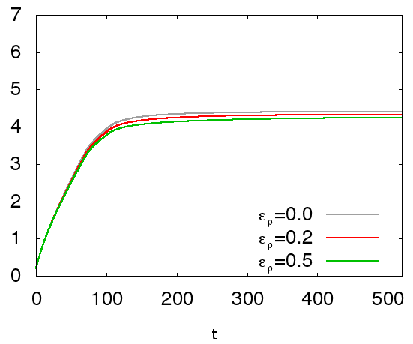}&
\hspace{-5mm}\includegraphics[width=4.7cm,height=4cm]{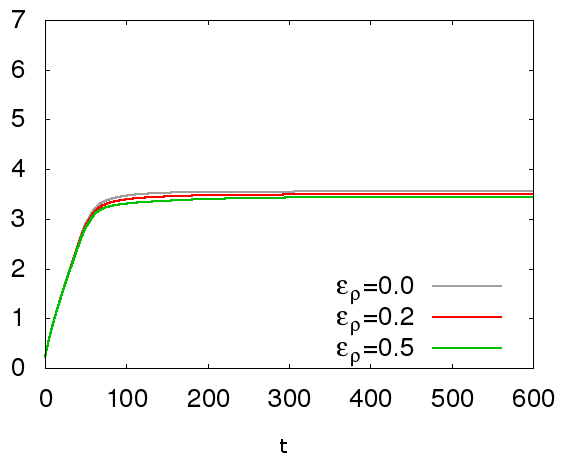}\\
\multicolumn{4}{c}{$a= 0.5$}\\ 
\hspace{-4mm}\includegraphics[width=4.7cm,height=4cm]{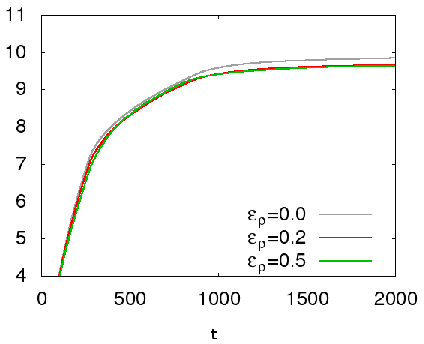}&
\hspace{-5mm}\includegraphics[width=4.7cm,height=4cm]{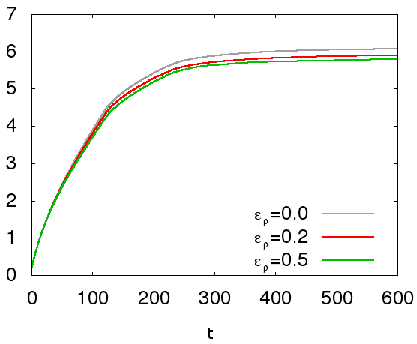}&
\hspace{-5mm}\includegraphics[width=4.7cm,height=4cm]{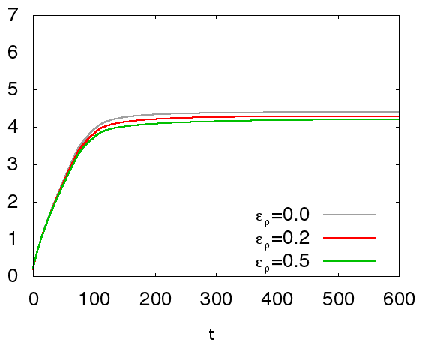}&
\hspace{-5mm}\includegraphics[width=4.7cm,height=4cm]{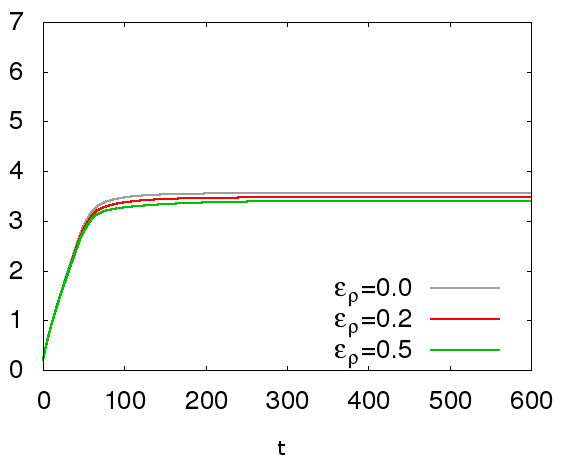}\\
\multicolumn{4}{c}{$a= 0.7$}\\ 
\hspace{-4mm}\includegraphics[width=4.7cm,height=4cm]{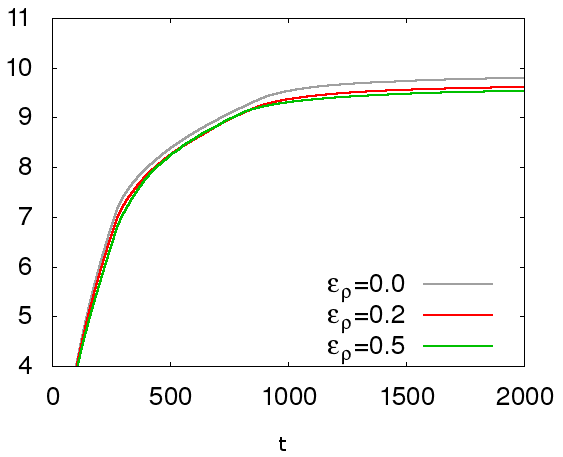}&
\hspace{-5mm}\includegraphics[width=4.7cm,height=4cm]{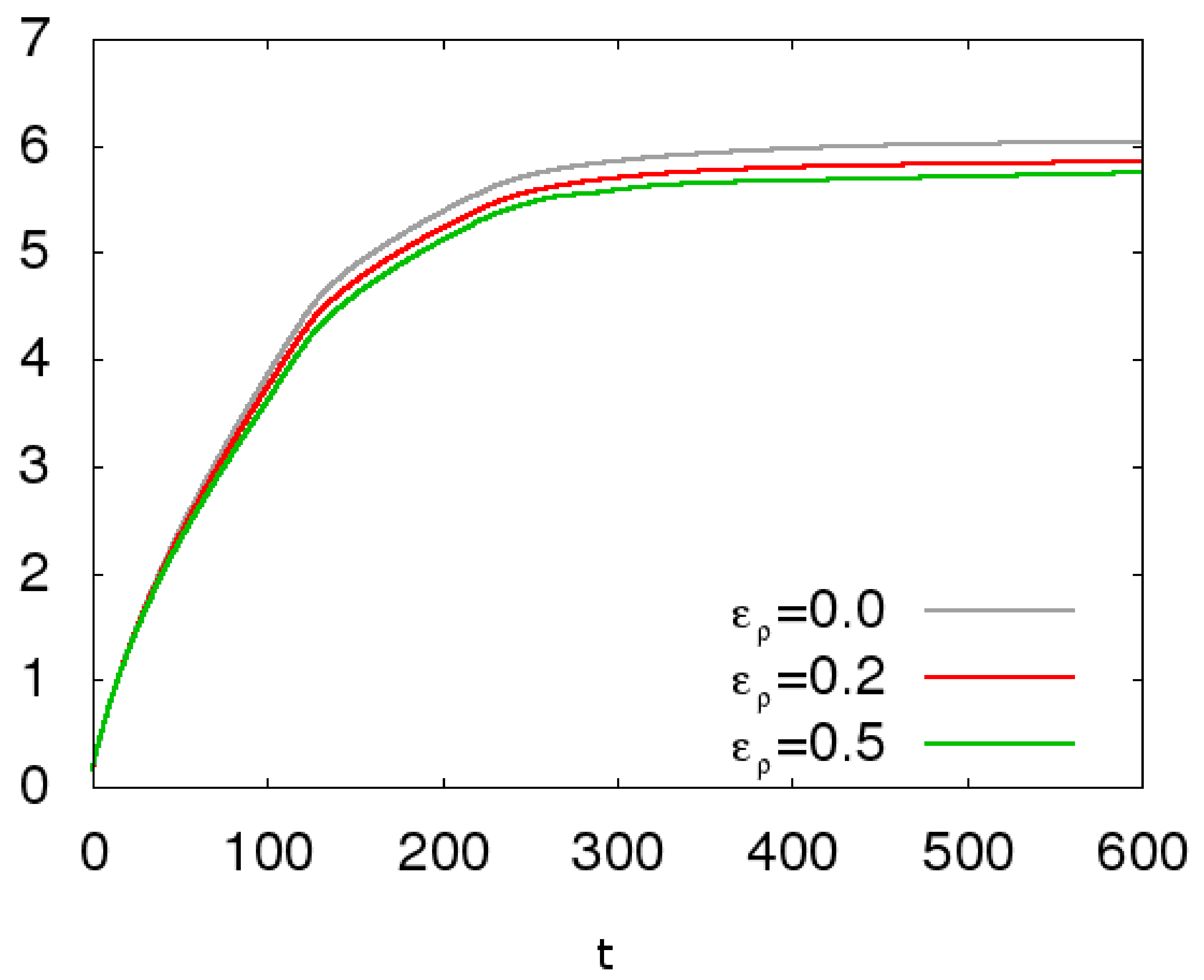}&
\hspace{-5mm}\includegraphics[width=4.7cm,height=4cm]{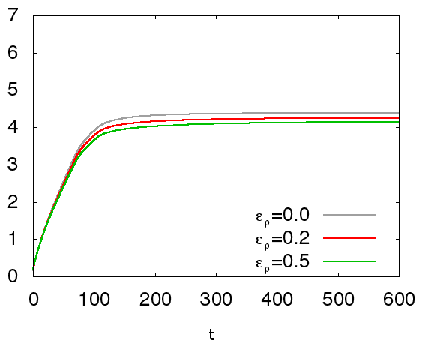}&
\hspace{-5mm}\includegraphics[width=4.7cm,height=4cm]{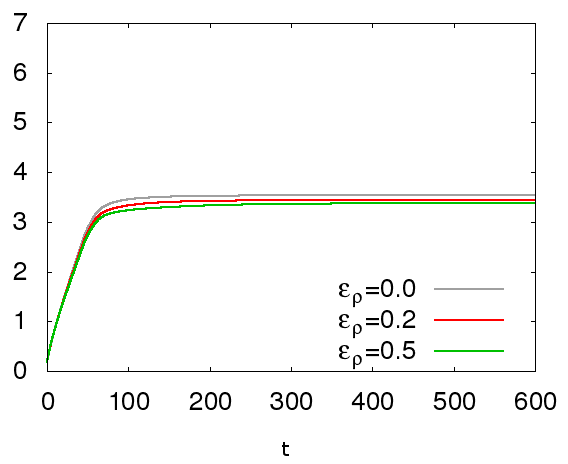}\\
\multicolumn{4}{c}{$a= 0.9$}\\ 
\hspace{-4mm}\includegraphics[width=4.7cm,height=4cm]{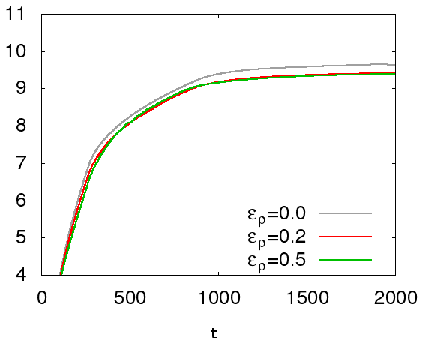}&
\hspace{-5mm}\includegraphics[width=4.7cm,height=4cm]{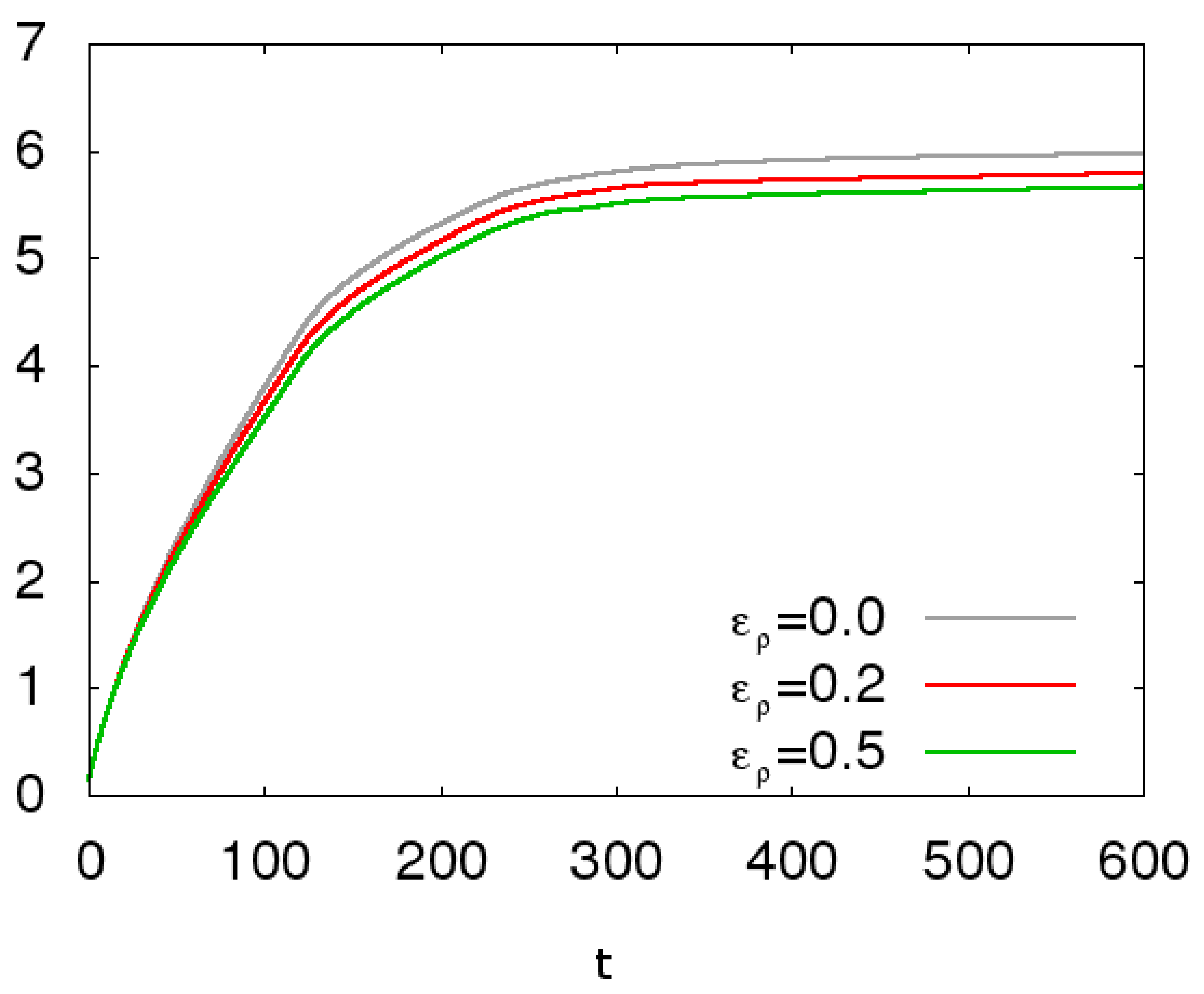}&
\hspace{-4.5mm}\includegraphics[width=4.7cm,height=4cm]{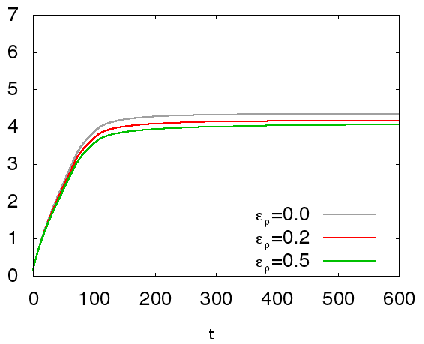}&
\hspace{-5mm}\includegraphics[width=4.7cm,height=4cm]{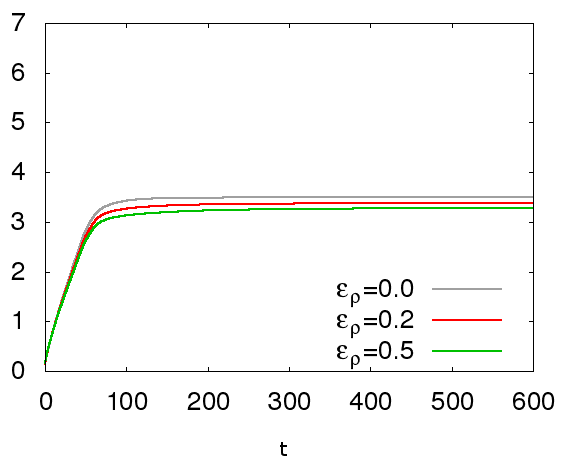}\\
\hline
\end{tabular}
  \caption{\label{MdotVsT} This figure shows the evolution of the mass accretion rates (the mass accretion rate is rescaled by $10^{-10}$).  
The mass accretion rate is measured near the event horizon of the BH (whose radius is a function of the spin parameter).  
Each model is run, at least, until steady state is acheived.
We show the cases considering various values of the Mach number ${\cal M}= 2,~3,~4,~5$ (columns) and different values for the BH spin parameter $a=0.3,~0.5,~0.7,~0.9$ (rows). 
In each figure we consider three values of the density gradient $\epsilon_{\rho}=0,~ 0.2,~0.5$.  We find that the mass accretion rate increases as the Mach number decreases.
In contrast, we can see that there is litle influence of the density gradient on the accretion rate.
We remind the reader that time is in units of $M$ with $1~\msun~\equiv~4.93~\times~10^{-6}$ s.}
\end{figure*}

\begin{figure*}
\begin{tabular}{cccc} \hline
 ${\cal M}= 2$ & ${\cal M}= 3$ & ${\cal M}= 4$ & ${\cal M}= 5$ \\ \hline
 && \\ 
\multicolumn{4}{c}{$a= 0$}\\ 
\hspace{-2mm}\includegraphics[width=4.7cm,height=4cm]{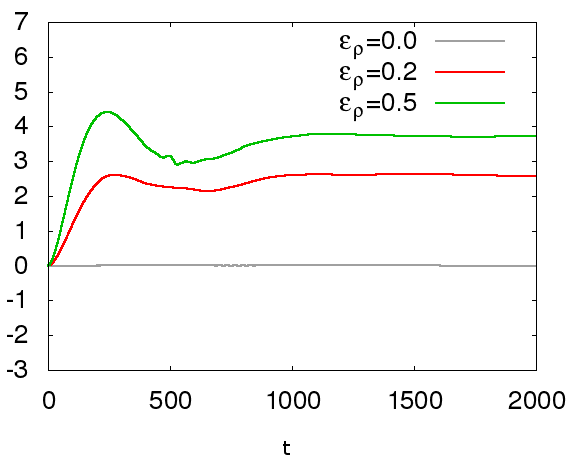}&
\hspace{-4mm}\includegraphics[width=4.7cm,height=4cm]{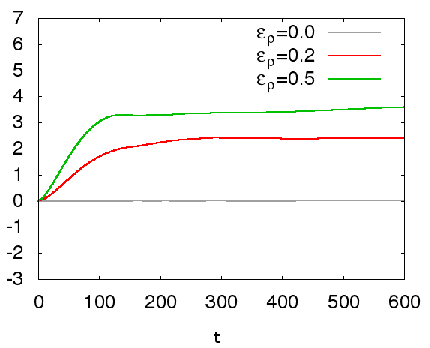}&
\hspace{-4mm}\includegraphics[width=4.7cm,height=4cm]{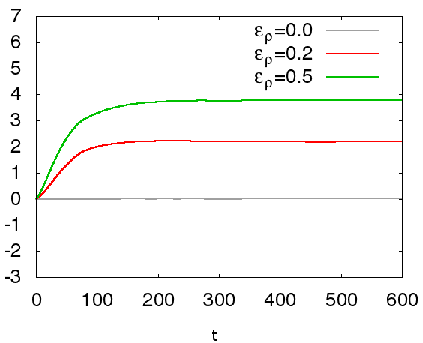}&
\hspace{-4mm}\includegraphics[width=4.7cm,height=4cm]{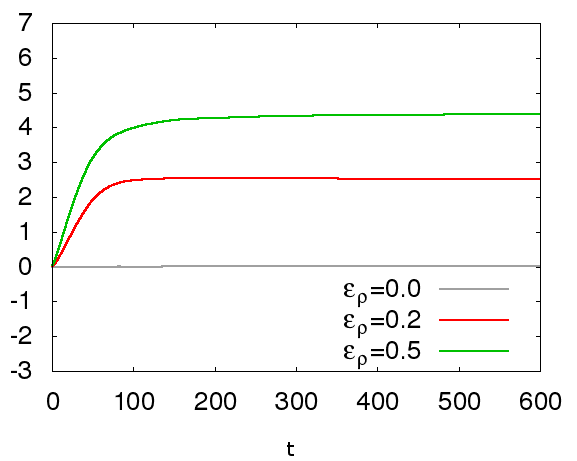}\\
\multicolumn{4}{c}{$a= 0.3$}\\ 
\hspace{-2mm}\includegraphics[width=4.7cm,height=4cm]{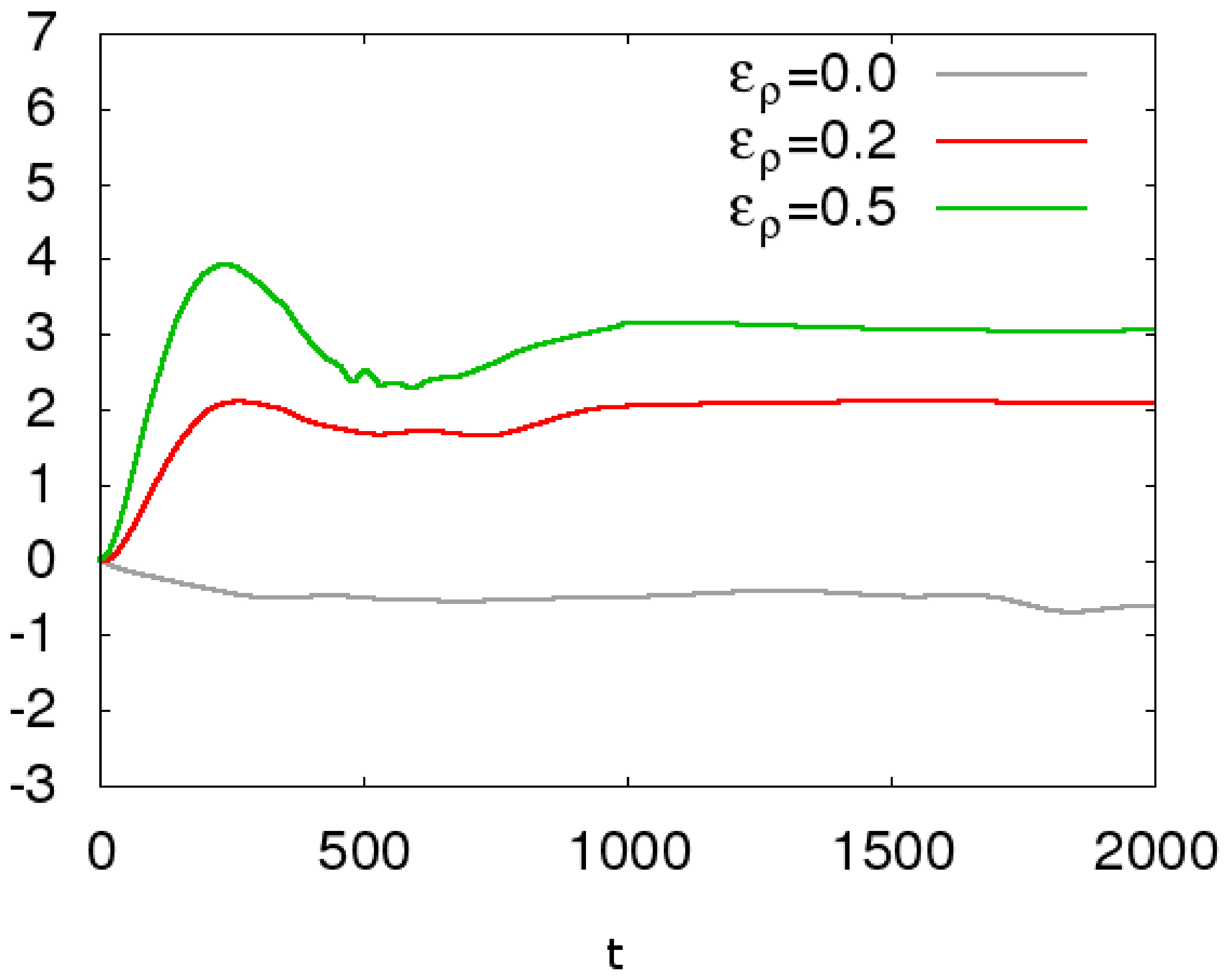}&
\hspace{-4mm}\includegraphics[width=4.7cm,height=4cm]{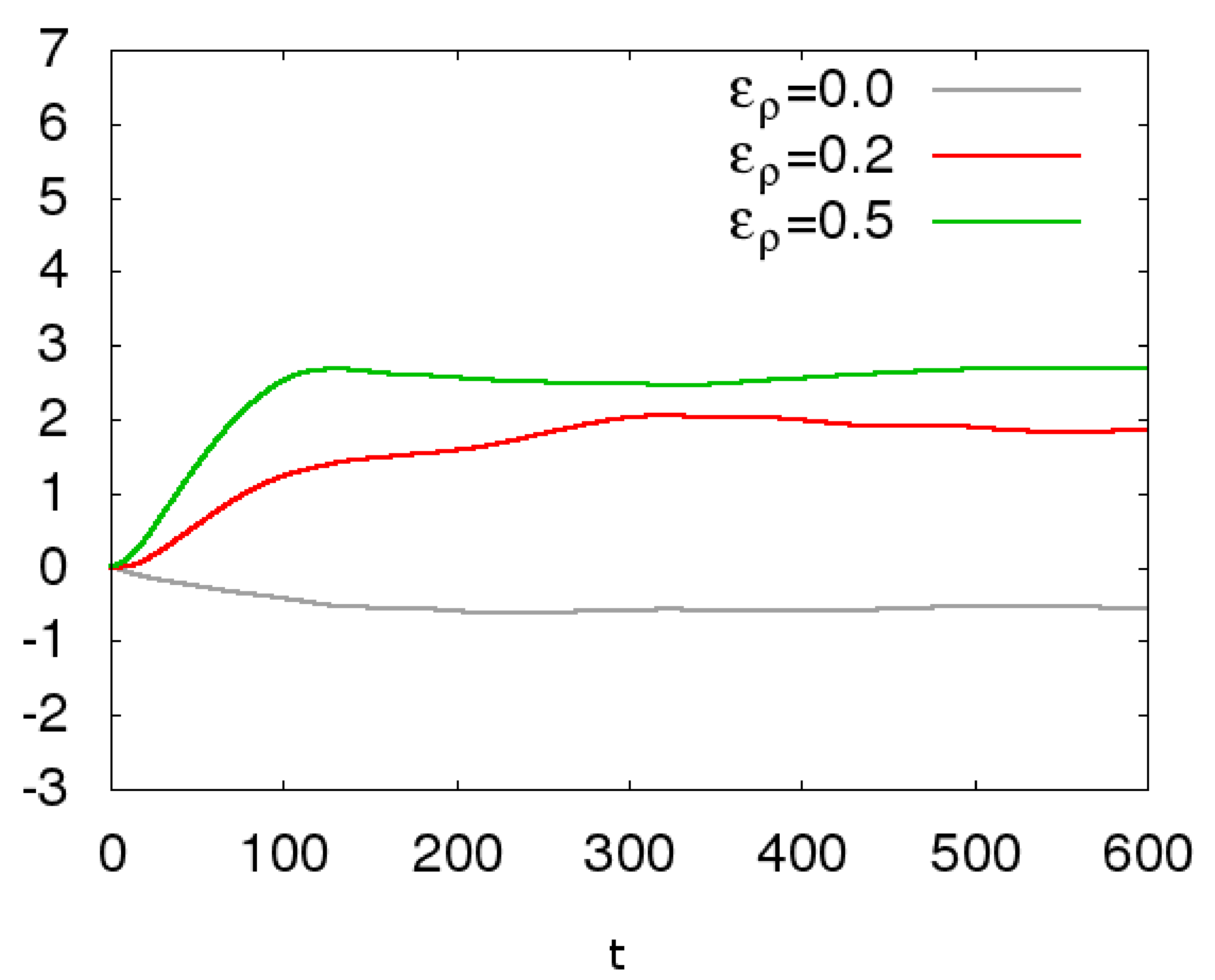}&
\hspace{-4mm}\includegraphics[width=4.7cm,height=4cm]{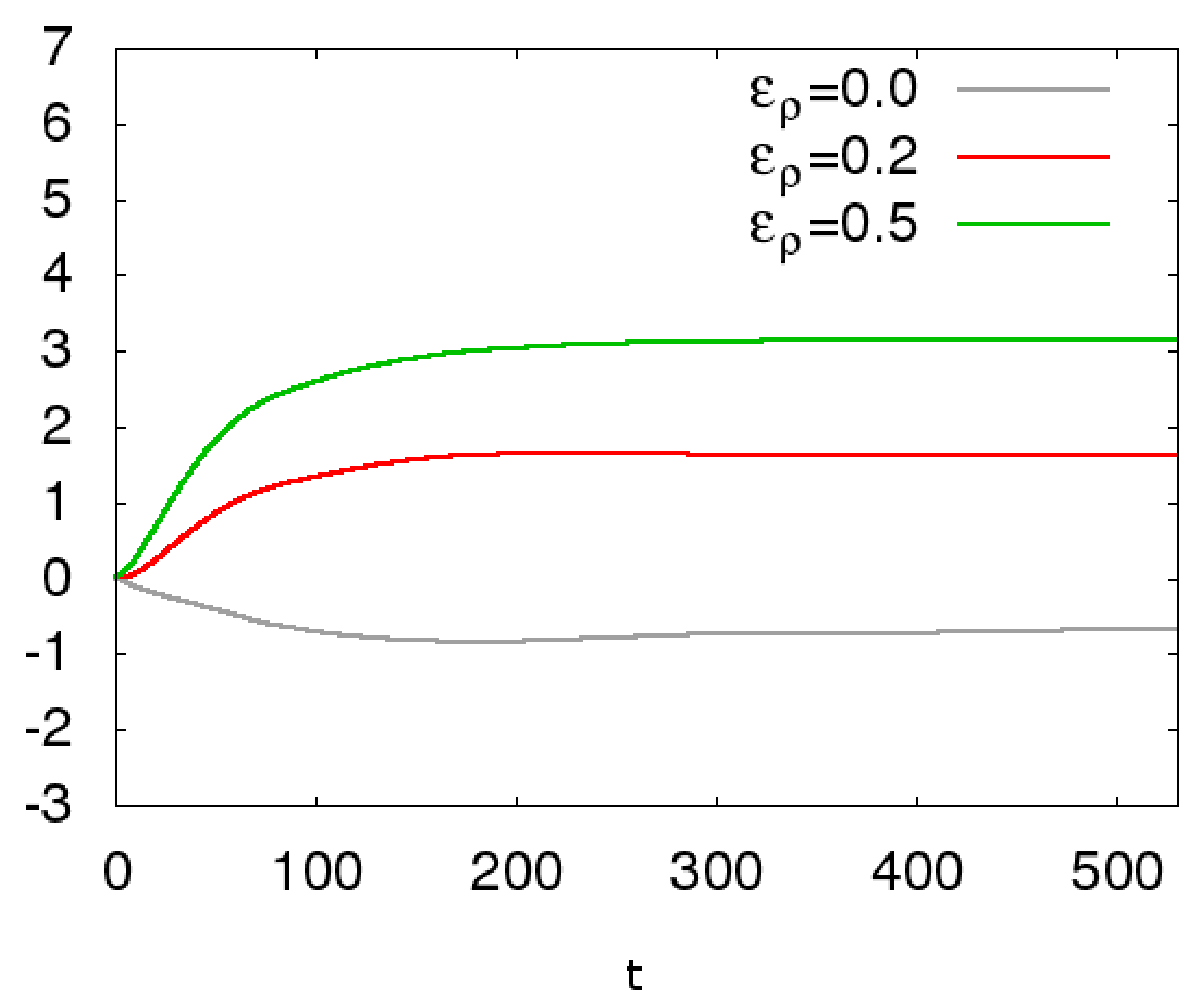}&
\hspace{-4mm}\includegraphics[width=4.7cm,height=4cm]{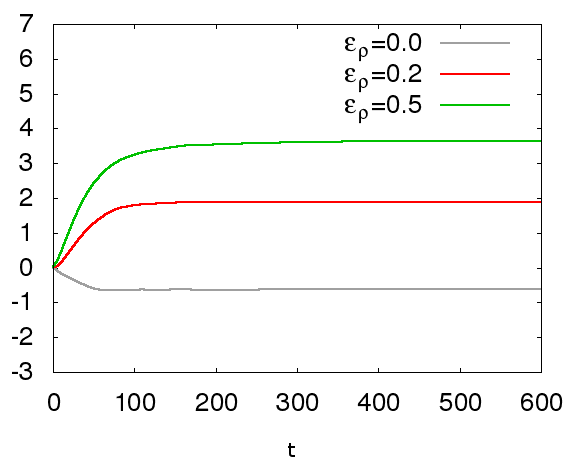}\\
\multicolumn{4}{c}{$a= 0.5$}\\ 
\hspace{-2mm}\includegraphics[width=4.7cm,height=4cm]{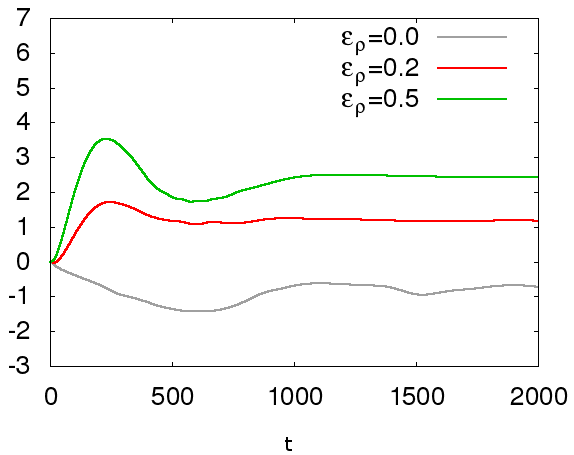}&
\hspace{-4mm}\includegraphics[width=4.7cm,height=4cm]{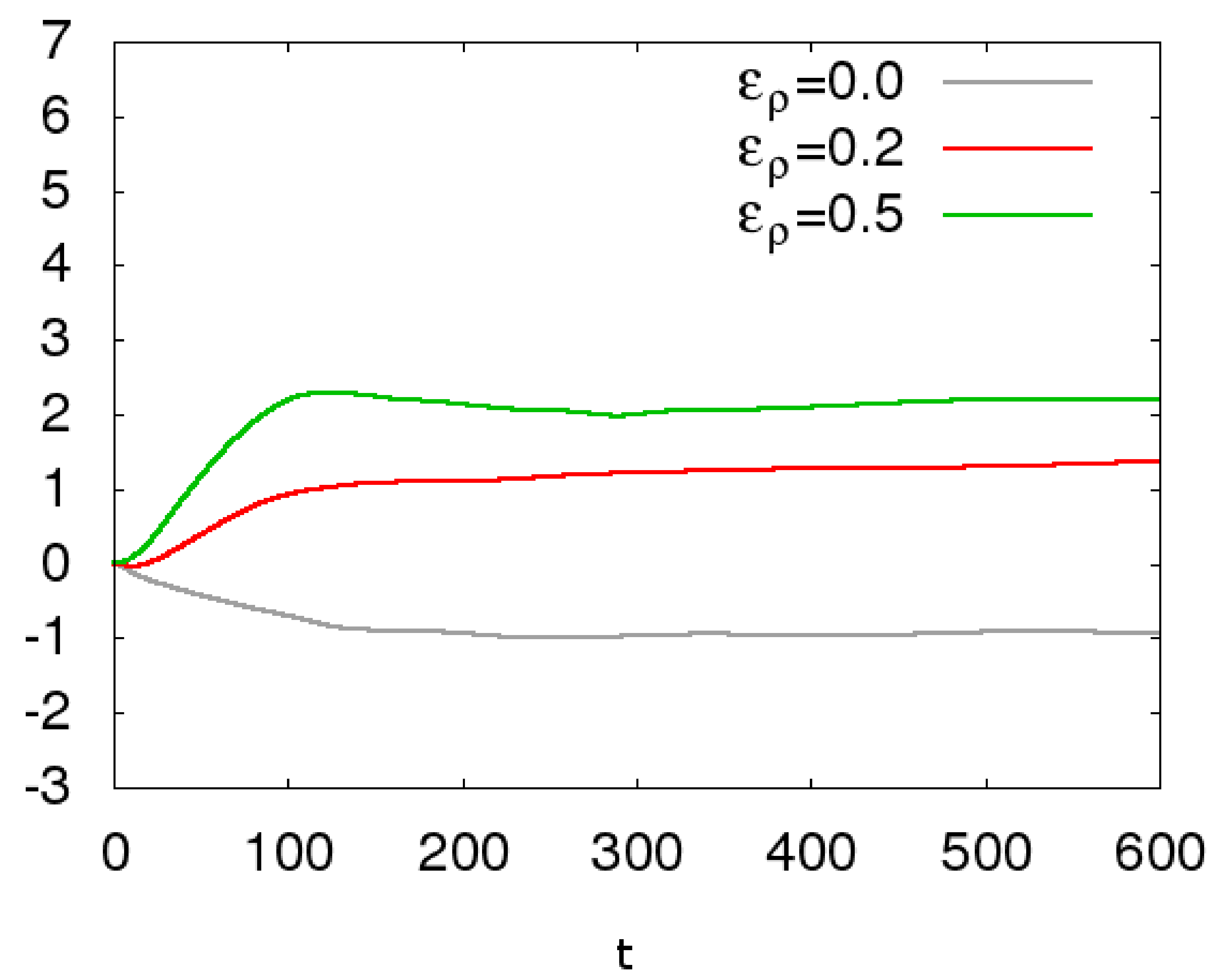}&
\hspace{-4mm}\includegraphics[width=4.7cm,height=4cm]{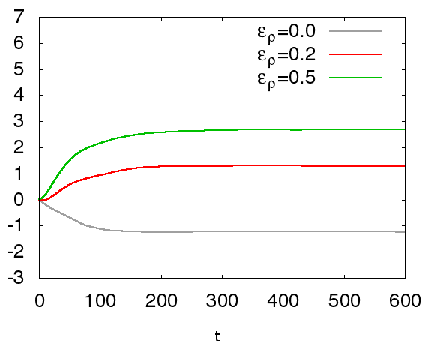}&
\hspace{-4mm}\includegraphics[width=4.7cm,height=4cm]{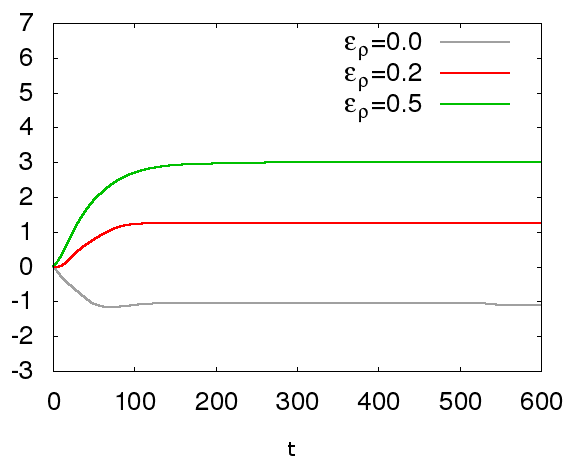}\\
\multicolumn{4}{c}{$a= 0.7$}\\ 
\hspace{-2mm}\includegraphics[width=4.7cm,height=4cm]{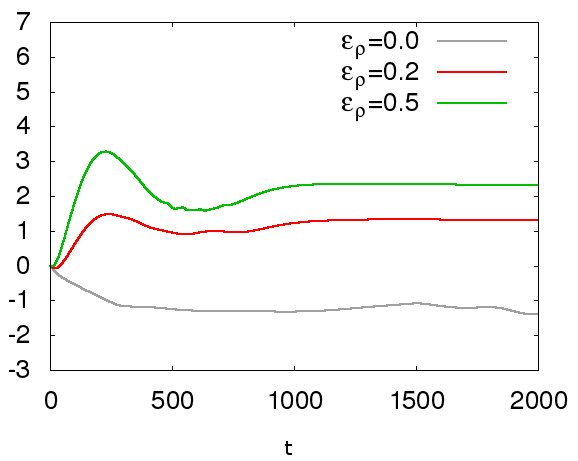}&
\hspace{-4mm}\includegraphics[width=4.7cm,height=4cm]{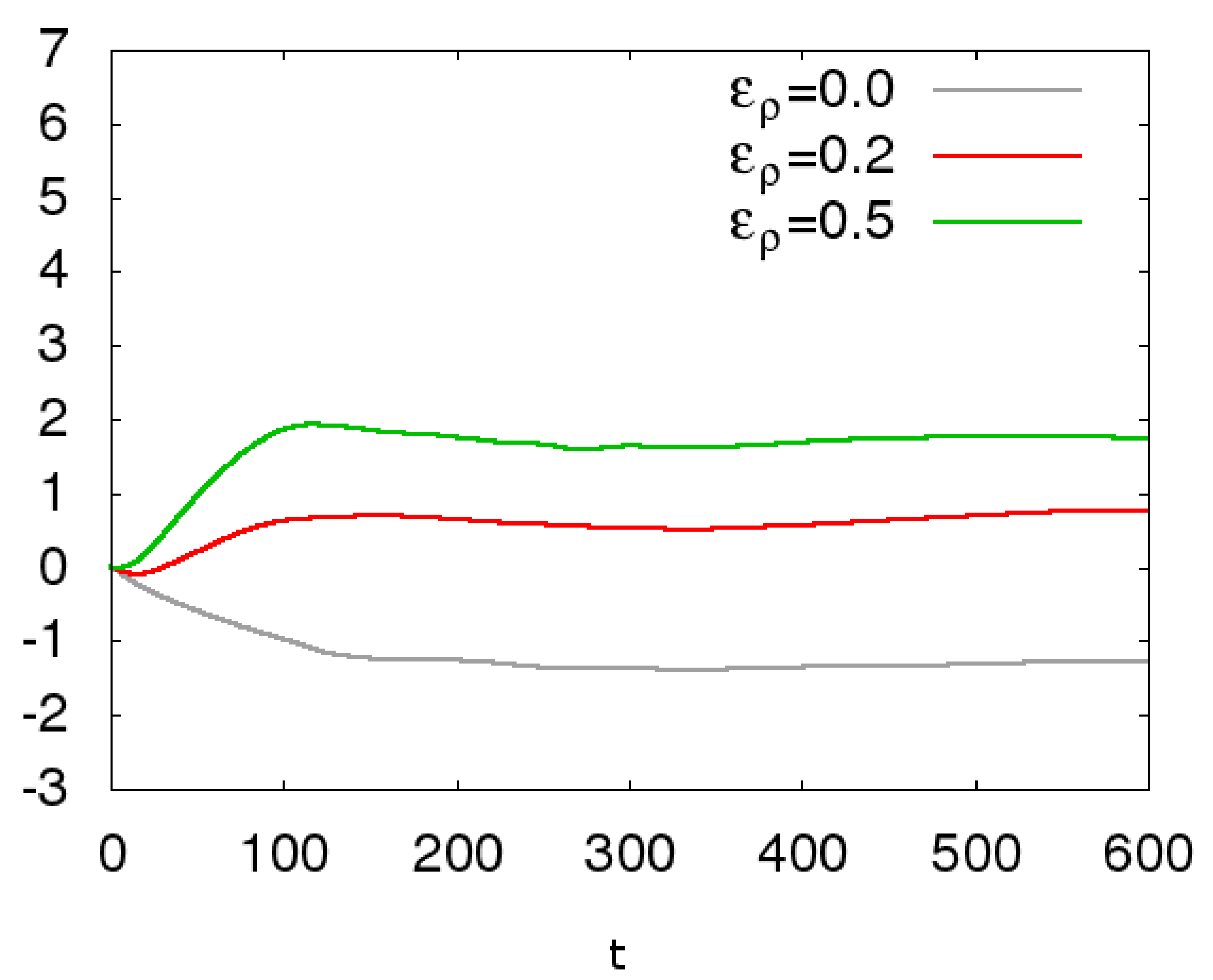}&
\hspace{-4mm}\includegraphics[width=4.7cm,height=4cm]{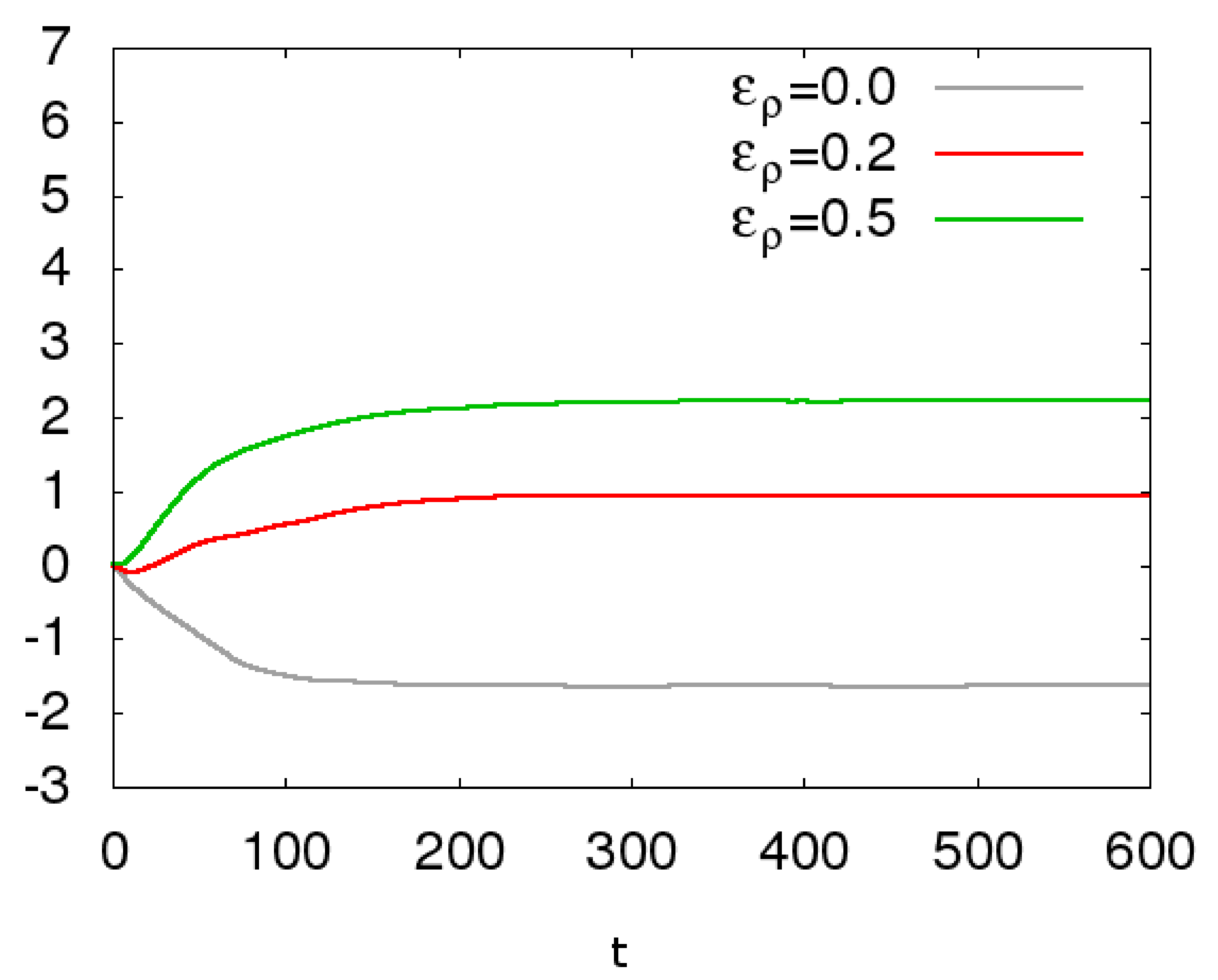}&
\hspace{-4mm}\includegraphics[width=4.7cm,height=4cm]{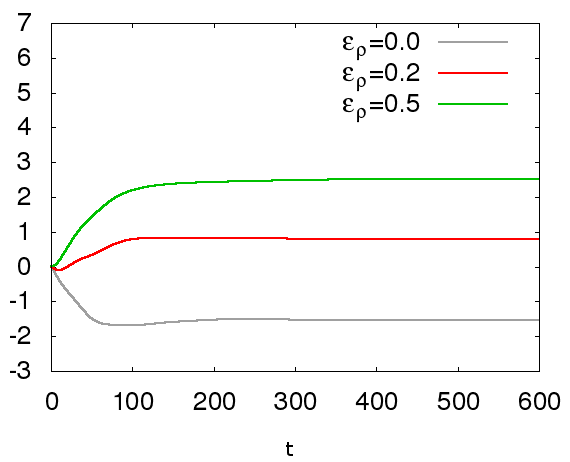}\\
\multicolumn{4}{c}{$a= 0.9$}\\ 
\hspace{-2mm}\includegraphics[width=4.7cm,height=4cm]{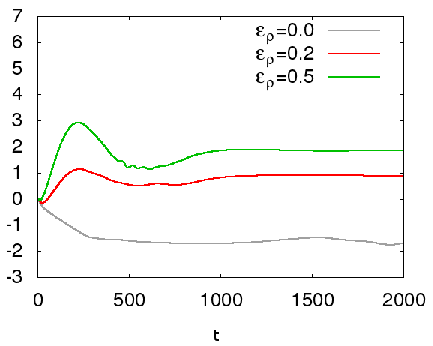}&
\hspace{-4mm}\includegraphics[width=4.7cm,height=4cm]{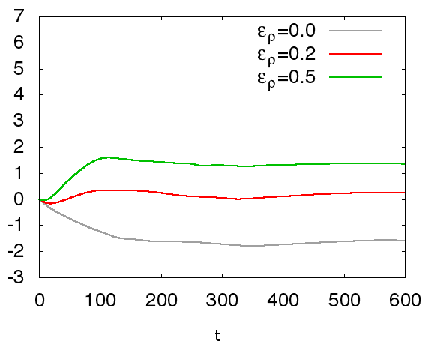}&
\hspace{-4mm}\includegraphics[width=4.7cm,height=4cm]{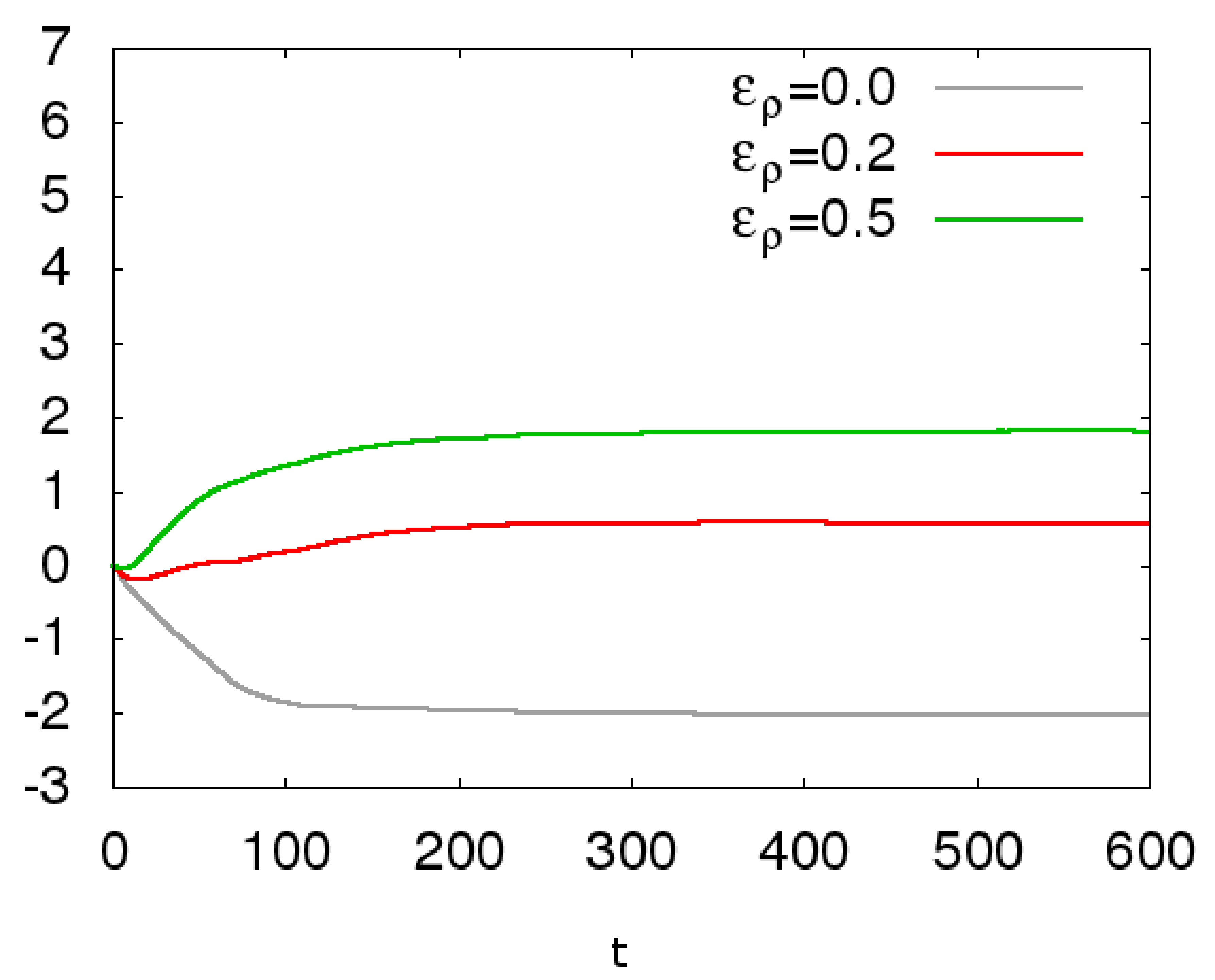}&
\hspace{-4mm}\includegraphics[width=4.7cm,height=4cm]{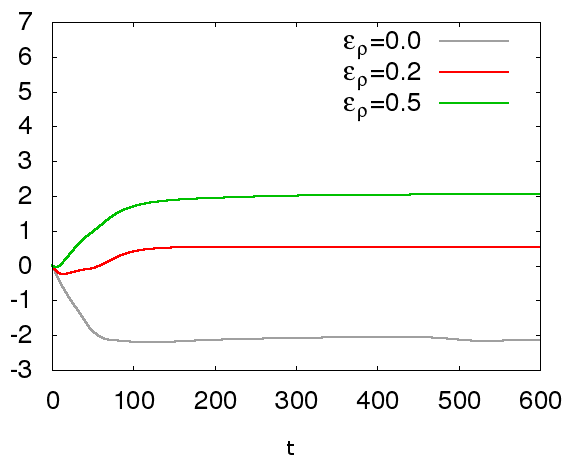}\\
\hline
\end{tabular}
  \caption{\label{PdotVsT} In this figure we show the angular momentum accretion rates vs time (the angular momentun rate is rescaled by $10^{-10}$). Again we we show the models corresponding to Mach numbers ${\cal M}= 2,~3,~4,~5$ (columns) and for angular parameter $a=0.3,~0.5,~0.7,~0.9$ (rows).  For each value of the Mach number we study three values of the density gradient $\epsilon_{\rho}=0,~ 0.2,~0.5$.  We can observe notable  changes in the angular momentum rates when the density gradient parameter increase. All rates are measured near the event horizon.  We remind the reader that time is in units of $M$ with $1~\msun~\equiv~4.93~\times~10^{-6}$ s.}
\end{figure*}

We have studied the parameter space for mass and angular-momentum accretion rates.
Our parameters are the Mach number, ${\cal{M}}$ (the columns in our plot array in figures~\ref{MdotVsT} and~\ref{PdotVsT}), the spin of the BH, $a$ (the rows in our plot arrays in figures~\ref{MdotVsT} and~\ref{PdotVsT}) and the density-gradient parameter of the gas,
$\epsilon_\rho$ (represented by the different color lines in our plot array in figures~\ref{MdotVsT} and~\ref{PdotVsT}).

All the shown figures (\ref{fig:Morph}) are snapshots of the system in steady state, or the stationary regime. 
These figures illustrate the BH rotating counter-clockwise for positive values of $a$ and with the flow moving from left to right.
That is, they correspond to timescales where mass and angular momentum accretion rates have stabilized and thus the curves (in figures~\ref{MdotVsT} and~\ref{PdotVsT}) have plateaued.

Our results, as seen in the first column (for $\epsilon_\rho = 0$) of figures~\ref{fig:Morph} correctly reproduce the expected properties of the shock cone when there exists no density gradient in the  medium where the black hole moves.
I.e., we observe the formation of a symmetric shock cone whose width depends on the Mach number; as the Mach number increases the shock-cone angle decreases \citep{1998ApJ...494..297F,1998ApJ...507L..67F,1998MNRAS.298..835F,1999MNRAS.305..920F,2012MNRAS.426..732C,2013MNRAS.429.3144L}.

\subsection{Morphology}\label{subsec:Morph}

\begin{figure*}
 \begin{tabular}{ccc} \hline
 $\epsilon_{\rho} = 0$ & $\epsilon_{\rho}=0.2$ & $\epsilon_\rho = 0.5$ \\ \hline
 && \\ 
\vspace{-5mm}\\
   \hspace{-10mm}\includegraphics[width=6.7cm]{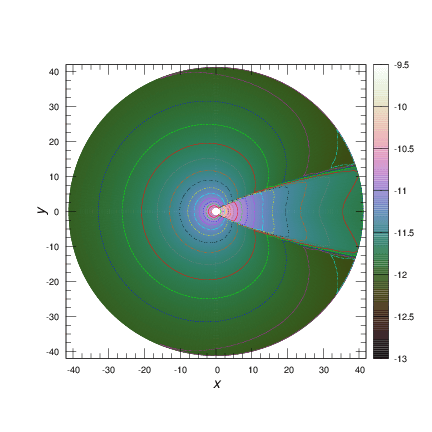}&
   \hspace{-9mm}\includegraphics[width=6.7cm]{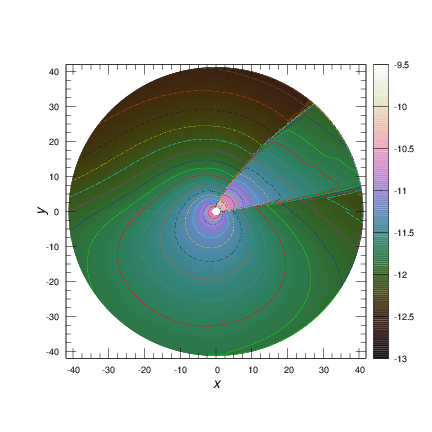}&
   \hspace{-9mm}\includegraphics[width=6.7cm]{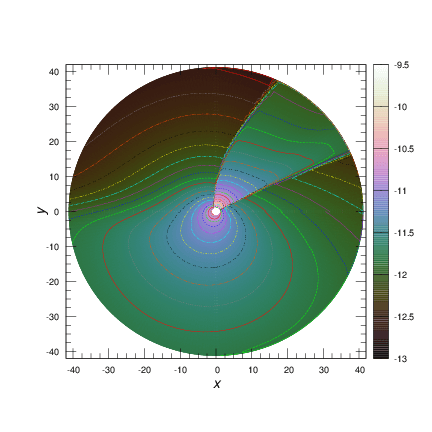}\\
\vspace{-15mm}\\
   \hspace{-10mm}\includegraphics[width=6.7cm]{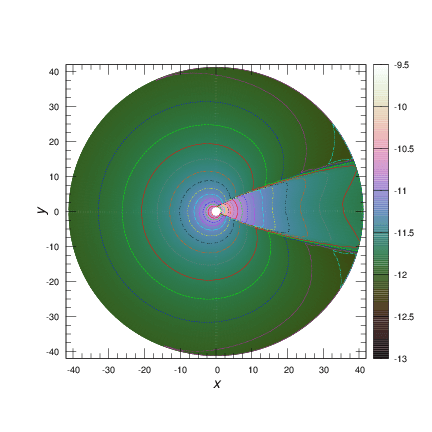}&
   \hspace{-9mm}\includegraphics[width=6.7cm]{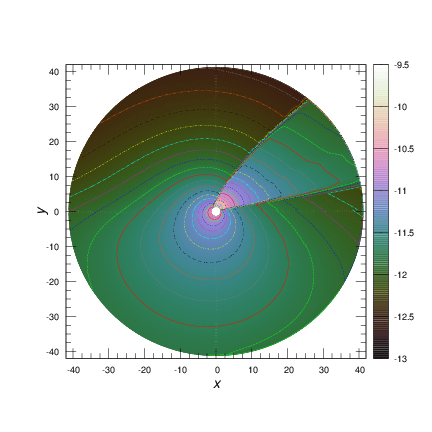}&
   \hspace{-9mm}\includegraphics[width=6.7cm]{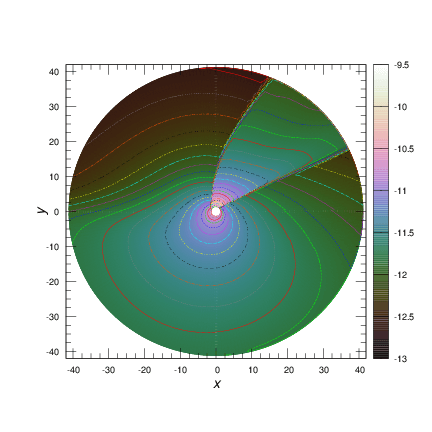}\\
\vspace{-15mm}\\
    \hspace{-10mm}\includegraphics[width=6.7cm]{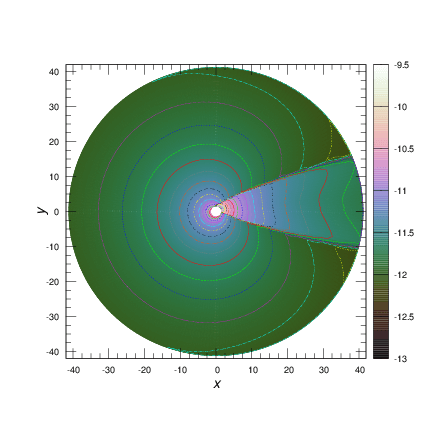}&
    \hspace{-9mm}\includegraphics[width=6.7cm]{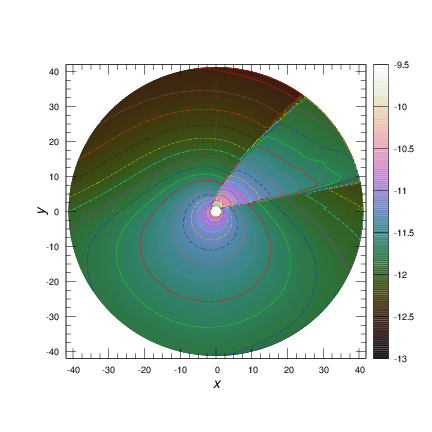}&
    \hspace{-9mm}\includegraphics[width=6.7cm]{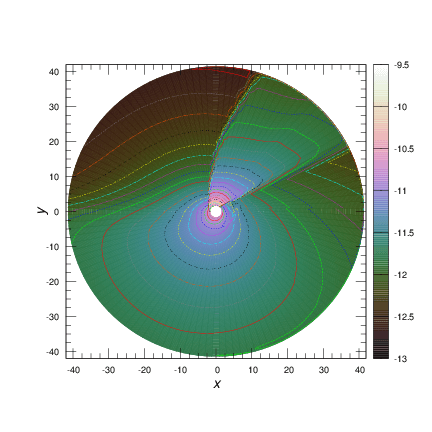}\\
\vspace{-8mm}\\
\hline
\end{tabular}
\caption{\label{fig:Morph} Morphology of the rest mass density at stationary state, we can observe the
shock cones are dragged due to the density gradients. We show models with Mach number ${\cal M}=~5$. Different rows, from top to bottom show results for $a=~0,~0.5,~0.9$. The columns, from left to right show density gradient parameter $\epsilon_{\rho} = 0, ~0.2,~0.5$. The contour lines range is $[-12.6,-10]$ and the space between a line and another is $0.1$ in each plot. We remind the reader that length is in units of $M$ with $1~\msun~\equiv~1.48~\times~10^{5}$ cm.}
\end{figure*}

We present, for the first time, the 2D morphology of the relativistic BHL accretion onto a BH considering density gradients (figures~\ref{fig:Morph}).
It is worth mentioning that in order to illustrate the general morphology of the system, we present only the case of ${\cal M} = 5$; however we have covered all configurations prented in this paper.
This first attempt is done in slab symmetry, which is the first of a series of steps towards more realistic 3D simmulations.
The color gradient represents the logarithm of the gas density in geometrical units normalized to the BH mass.
The contour lines simply emphasize the density gradient.
These figures dramatically illustrate the effect that the density gradient has on the shock cone once steady state is acheived.

From the first column of figures \ref{fig:Morph} we observe that \citep[as in][ where zero density gradient is considered]{1999MNRAS.305..920F} as $a$ increases so does the induced angular momentum in the wind.
Now, as the density gradient increases, we observe the most notable feature in these figures, i.e., the shock cone is further pushed towards the lower density region.

\subsection{Mass and Angular Momentum Accretion Rates}\label{subsec:Acc}

From the plots on figures~\ref{MdotVsT} and~\ref{PdotVsT}, we observe that, as expected, the mass accretion rate decreases as the Mach number, ${\cal M}_{\infty}$, increases.
Unlike the Newtonian case where no trend between the mass-accretion-rate fluctuations and the density gradient was discerned \citep{1999A&A...346..861R} our results exhibit a stationary state rather quickly and show that as $\epsilon_\rho$ increases the mass accretion rate slightly decrases.
The faster the fluid moves with respect to the BH the faster the steady states settle in.
The mass accretion rate is, mostly, independent of the spin parameter of the BH.
The reason for the the steady state settling is rather quickly may be due to the fact that our simmulations are done in slab symmetry as opposed to 3D.
It could also be that we are dealing with relativistic flows.
We confirm that the accretion rates do decrease slightly when the density gradient increases.
Besides, when the BH spin increases the effects of the wind density gradient on the mass accretion rate become slightly stronger.

We further explore cases where the accretor has angular momentum.
As ${\cal M}_{\infty}$ decreases the difference between the mass accretion rate, for different $\epsilon_\rho$, becomes larger.  
This is not noticeable, however, in the case where ${\cal M}_{\infty} = 2$, because for lower velocities steady state settles in much later (notice the timescale is $\sim 3$ times longer and it is still not fully achieved; this would be consistent with the Newtonian results in \citealt{1999A&A...346..861R}).

The behavior of the angular momentum accretion rate (figs.~\ref{PdotVsT}), instead, clearly shows that for higher wind velocities (increasing ${\cal M}_{\infty}$) steady state is reached much more quickly.  
For ${\cal M}_{\infty} = 4$ and ${\cal M}_{\infty} = 5$ steady state is achieved after about $t = 200$  and $t = 100$, respectively, regardless of BH spin or density gradient.
Instead, looking at ${\cal M}_{\infty} = 3$ and, espacially at ${\cal M}_{\infty} = 2$ it is clear that a steady state is not fully reached in $t = 600$ for the former and not even in $t = 2000$ for the later.  
This trend is more noticeable for lower density gradient and/or higher BH spin.
In other words, it appears that increasing the density gradient helps stabilize the angular momentum accretion rate at an earlier time.
We also observe that the angular momentum rates extered on the wind decrease into negative values further and further the more the BH spin, $a$, increases.  
It can also be seen that higher Mach number has the effect of slightly increasing the angular momentum accretion rate at steady state for large density gradient ($\epsilon_\rho = 0.5$).  For lower density gradient ($\epsilon_\rho = 0.2$) this trend is not evident.
And, for no density gradient ($\epsilon_\rho = 0$) there seems to exist no correlation.


We have produced, as well, a set of simulations where the spin of the BH counterrotates, $a<0$.  
The first case has $a=-0.9$, $\gamma = 5/3$ and ${\cal M}_{\infty}= 0.3$; the second case has $a=-0.9$, $\gamma = 5/3$ and ${\cal M}_{\infty}= 0.4$; and, finally the third case has $a=-0.99$, $\gamma = 5/3$ and ${\cal M}_{\infty}= 0.4$.
In all three cases we performed simulations for the same three values of the density gradient, i.e., $\epsilon_\rho = 0$, 0.2 and 0.5.
Opposite to the cases described above, here the angular momentum of the accreted material is parallel to that of the BH, thus, it is expected that mass accretion will tend to increase the spin of the BH on the long run.

As can be observed from the simulations for low Mach number (see fig.~\ref{a05_v01}), a disk-like structure forms.
The velocity field has positive radial coordinate, thus, little or no accretion occurs.

\section{Discussion and conclusions}
\label{sec:conclusions}

We have performed a parameter study of BHL accretion of a relativistic wind with density gradient onto Schwarschild and Kerr BHs with different spin parameters.  A discussion of our results follows in the next paragraphs.

\begin{figure*}
 \begin{tabular}{ccc} \hline
 ${\cal M}_{\infty}= 3$, $a=-0.9$ & ${\cal M}_{\infty}= 4$, $a=-0.9$ & ${\cal M}_{\infty}= 4$, $a=-0.99$ \\ \hline
 && \\ 
  \includegraphics[width=5.5cm]{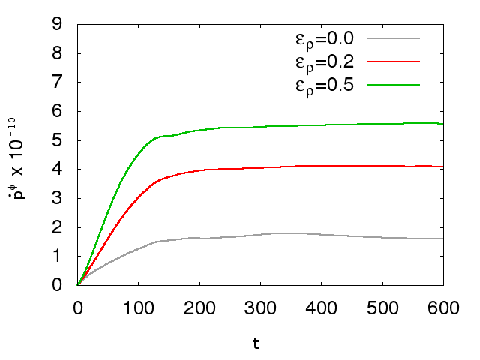}&
  \includegraphics[width=5.5cm]{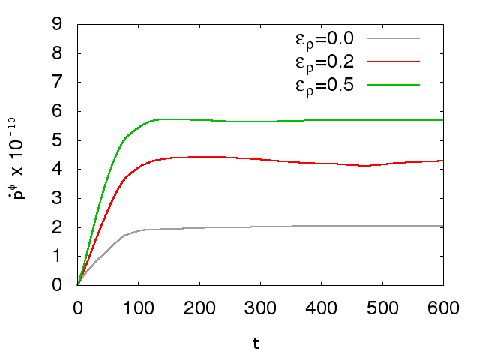}&
  \includegraphics[width=5.5cm]{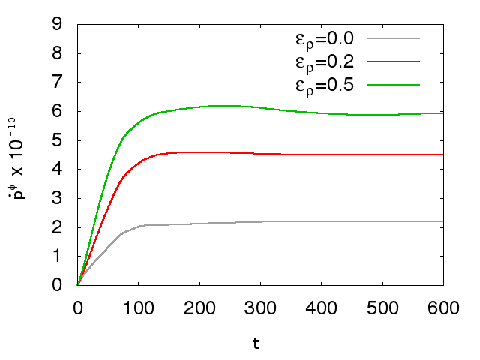}\\\vspace{-2mm}
  \end{tabular}
  \caption{\label{a09v03g53} In this figure we show the angular momentum rates of the special cases considering counterrotating black holes and density gradients $\epsilon_\rho=0, ~0.2,~0.5$. In contrast with positive values  of the angular momentum of the black hole we can observe that, in all cases, the angular momentum rates are positive.}
  \end{figure*}

Comparing figures~\ref{MdotVsT} vs~\ref{PdotVsT}, and~\ref{a09v03g53}, it is interesting to note that the mass accretion rate, unlike the angular momentum rate (see first column plots in fig.~\ref{PdotVsT}), does not depend strongly on the wind density gradient regardless of Mach number or BH spin.
As can be observed from our fig.~\ref{MdotVsT}, observing the plots from left to right, as the Mach number increases the mass accretion rate decreases; in agreement with previous studies as well as theory.
It is also important to note that, as the Mach number increases, steady state is obtained much more rapidly.  
This is further implied by the results in \citet{1999A&A...346..861R}, where only Newtonian velocities are acheived and the plots show much more variabilty in both mass and angular momentum accretion.

From the plots in figure~\ref{PdotVsT} we find that the wind may be accreted with low or even negative values of angular momentum with respect to the spin of the BH.
This implies that, were a significant amount of material to be accreted (over long periods of time), the spin of the BH could be brought down significantly or even reversed in the cases where the angular momentum has negative values.
On the other end of the BH spin spectrum, from fig.~\ref{a09v03g53}, it is clear that the spin of a BH can increase over time if the spin and the wind density gradient are properly aligned.

From the morphology plots (figs.~\ref{fig:Morph}), we observe a new signature, i.e., the Mach cones wrap around the BH, even with low or nil spin, due to the density gradient of the winds.

\begin{figure*}
 \begin{tabular}{cc} 
 \includegraphics[width=10cm]{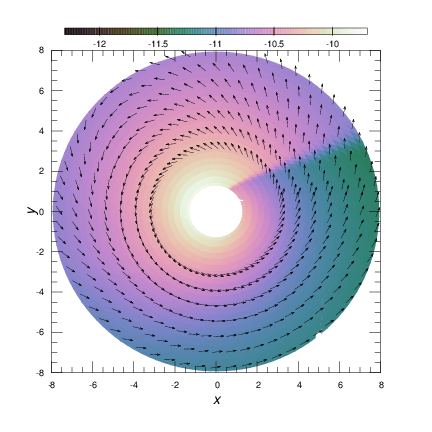}&\hspace{-15mm}
 \includegraphics[width=10cm]{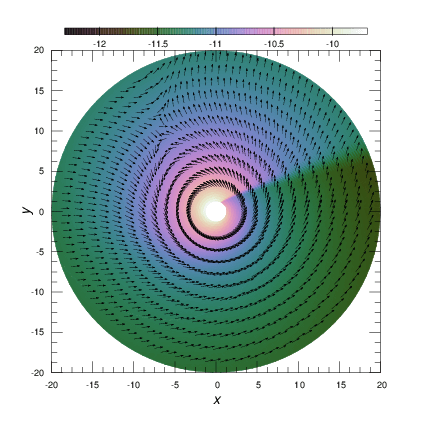}\vspace{-10mm}\\
 \vspace{-2mm}
 \hspace{-7mm}
 \includegraphics[width=9cm]{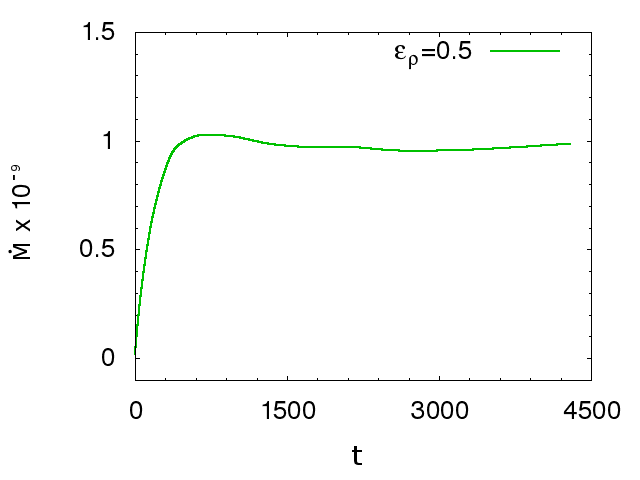}&\hspace{-25mm}
\includegraphics[width=9cm]{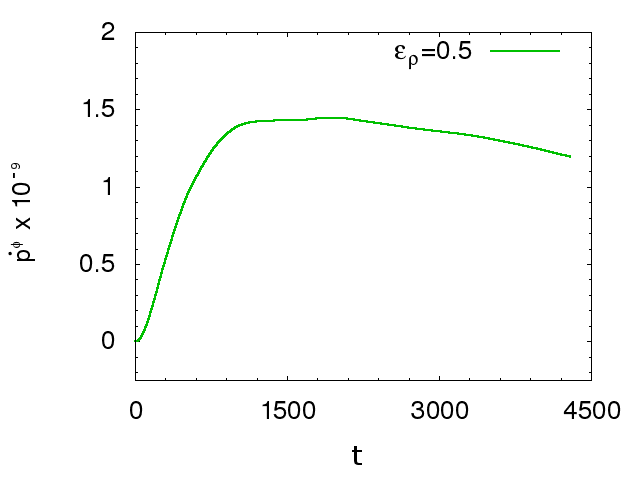}\\
  \end{tabular}
  \caption{\label{a05_v01} We show, on the top two figures, the morphology of the rest mass density and the velocity field vectors. We show two close-ups, one at $r=8$ (left) and other at $r=20$ (right). On bottom we show the evolution of mass (left) and angular momentum (right) accretion rates. This special case corresponds to a model with paramenters $\gamma = 5/3$, ${\cal M}_{\infty}= 1$, $a=0.5$ and $\epsilon_{\rho}=0.5$. In this model we find  the formation of a disk-like structure (see the 2D figures in the upper panels), the velocity-field vectors shows cuasi-circular trajectories of of the fluid elements. Notice that steady state has not been fully acheived for these cases after 4,500 $M$.}
\end{figure*}

Probably, due to the fact that our runs do stabilize rather quickly, we observe that an accretion disk starts to form for $\gamma = 5/3$, low Mach number, high BH spin and high density gradient ($\epsilon_\rho = 0.5$; see Fig. \ref{a05_v01}).  

As noted above, in the simulations where a disk forms the velocity field has positive radial component, which would imply no accretion occurs.
However, the fluid modeled by our code has no account of radiative losses. 
Thus, in a more realistic simulation it is highly likely that energy would be radiated away and angular momentum would be transported outwards within the disk allowing for substantial accretion onto the BH.

\subsection{Astrophysical Scenarios and Applications}

From figure~\ref{MdotVsT} we can observe that for a $\sim 10 \msun$ BH steady-state accretion can be acheived after a few milliseconds. 
For IMBHs the timescales necessary for reaching steady-state accretion go from seconds to hours.
Whereas for SMBHs said timescales go from hours to years.

In the hypernova-explosion scenario for sBHs and IMBHs, accretion will probably last a few minutes thus little mass or angular momentum may be accreted.

For the scenario where a SMBH crosses through the jet of a companion SMBH substantial mass may be accreted if the orbit and the supply of material to the BH producing the jet are stable.  However, the effects of the density gradient modifying the spin of the BH may be cancelled out if the axis of the jet is on the orbital plane.
This occurs because as the BH enters the cone of the jet the density increases, however, as the BH exits the jet, the opposite effect takes place, thus cancelling most of the effect. 
If the axis of the jet is slightly off the orbital plane then there is a component of density gradient that does not get cancelled out and a BH spin can be built up over many orbital periods.  
If two jets, as opposed to one, exist and they are symmetric, the effect of the density gradient on the BH spin will also be cancelled out.
Were the jet to precess (so that the axis of the jet is sometimes above the orbital plane and sometimes below) on a timescale much larger than the orbital period spin reversal could be observed on the BH that cuts through the jet.  
Maybe more important is the fact that these processes are accreting mass
but more likely than not, no angular momentum, thus the spin, $a$, of the BH will decrease (as the spin $a \propto J/M^2$).

The reader may ask how likely it is that the jet of a SMBH may hit another one.  
There are a few examples where binary SMBHs have been observed \citep[e.g.,][]{2008Natur.452..851V,2011Natur.477..431F,
2012MNRAS.427...77V,2015Natur.518...74G}.
The most relevant parameters to estimate the likelihood of such an event, we suspect, are the following:
First, the jet must have a large angle, otherwise chances are really small.
Second, it is suspected 

that these binaries are the product of a merger of two galaxies, hence it is likely that if the spin of the SMBHs are similar to those of their hosts \citep[e.g.,][]{2012MNRAS.423.2533B} and they collide at random angles, the SMBHs spins will also be random.
Third, the impact parameter of the collision should, preferably, be small such that the BHs do not have to travel far within the newly merged galaxy, otherwise they may accrete substantial amounts of matter \citep[see, e.g.,][]{2002ApJ...567L...9A,2009MNRAS.396.1640D} which will have a prefered angular momentum (that of the host galaxy) and may reorient the axes towards that of the reshaped host \citep{2009MNRAS.396.1640D}, and, thus, preventing them from hitting each other with their jets.
The mass of the SMBHs will be important as well, as the spin of a more massive BH will be less affected by accretion as it drifts inwards to meet the companion BH \citep{2012MNRAS.423.2533B}.

\section*{Acknowledgments}
F.D.L-C gratefully acknowledges DGAPA postdoctoral grant to UNAM
 and financial support from CONACyT 57585. ACO had support from a DGAPA postdoctoral grant to UNAM. 
EMM had support from a CONACyT fellowship.  
This research has made use of NASA’s Astrophysics Data System as well as arXiv. 
No BHs were harmed during the process of this research (they were all properly fed).


\end{document}